\documentclass[aps,nofootinbib,noshowpacs,showkeys,preprintnumbers,amsmath,amssymb]{revtex4}
\usepackage{graphicx,color}
\usepackage{graphics}
\usepackage{rotating}
\usepackage{amssymb}

\usepackage{color}

\usepackage{epsfig}
\usepackage{dcolumn}
\usepackage{bm}
\usepackage{multirow}

\def\be{\begin{equation}}
\def\ee{\end{equation}}
\def\ba{\begin{eqnarray}}
\def\ea{\end{eqnarray}}
\def\bea{\begin{eqnarray}}
\def\eea{\end{eqnarray}}
\def\bes{\begin{subequations}}
\def\ees{\end{subequations}}
\def\bear{\begin{array}}
\def\eear{\end{array}}

\newcommand{\A}{{\mathcal{A}}}
\newcommand{\tA}{{\widetilde {\mathcal{A}}}}
\newcommand{\ta}{{\widetilde a}}

\newcommand{\tk}{{\widetilde k}}

\newcommand{\MSbar}{\overline{\rm MS}}  
  
\newcommand{\ovl}{\overline}

\begin{document}
\preprint{USM-TH-357}

\title{Bjorken polarized sum rule and infrared-safe QCD couplings}
\author{C\'esar Ayala$^{1}$}
\author{Gorazd Cveti\v{c}$^1$} 
\author{Anatoly V.~Kotikov$^2$}
\author{Binur G.~Shaikhatdenov$^2$}
\affiliation{$^1$Department of Physics, Universidad T{\'e}cnica Federico
Santa Mar{\'\i}a (UTFSM), Casilla 110-V, Valpara{\'\i}so, Chile\\
$^2$Joint Institute for Nuclear Research, $141980$ Dubna, Russia}

\date{\today}

\begin{abstract}
Experimental data obtained for the polarized Bjorken sum rule (BSR) $\Gamma_1^{p-n}(Q^2)$ are fitted by using predictions derived within a truncated operator product expansion (OPE) approach to QCD. Four QCD versions are considered: perturbative QCD (pQCD) in the ${\overline{\rm MS}}$ scheme, Analytic Perturbation Theory (APT), and 2$\delta$ and 3$\delta$ analytic QCD versions. In contrast to pQCD, these QCD variants do not have Landau singularities at low positive $Q^2$, which facilitates the fitting procedure significantly. The fitting procedure is applied first to the experimental data of the inelastic part of BSR, and the known elastic contributions are added after the fitting. In general, when 2$\delta$ and 3$\delta$ QCD coupling is used the fitted curves give the best results, within the $Q^2$-range of the fit as well as in extended $Q^2$-intervals. When the fitting procedure is applied to the total BSR, i.e., to the sum of the experimental data and the elastic contribution, the quality of the results deteriorates significantly.
\end{abstract}
\maketitle

\section{Introduction}
\label{sec:intr}

The polarized Bjorken sum rule (BSR) $\Gamma_1^{p-n}(Q^2)$ \cite{BjorkenSR} is an important spacelike QCD observable for various reasons. It is a difference of the first moment of the spin-dependent structure functions of proton and neutron, therefore its isovector nature makes it easier to describe it theoretically, in pQCD in terms of OPE, than the separate integrals of the two nucleons. Further, high quality experimental results for this quantity, obtained in polarized deep inelastic scattering (DIS), are now available in a large range of spacelike squared momenta $-q^2 \equiv Q^2$: $0.054 \ {\rm GeV}^2 \leq Q^2 < 5 \ {\rm GeV}^2$  \cite{EG1a,EG1b,DVCS,E143,COMP1,COMP2,E155,SMC,HERMES}. In particular, the newer experimental results \cite{DVCS} with highly reduced statistical uncertainties, extracted mainly from the Jefferson Lab CLAS EG1-DVCS experiment \cite{EG1DVCS} taken on polarized targets made of protons and deuterons, render BSR an attractive quantity to test on it various extensions of pQCD to low $Q^2 \lesssim 1 \ {\rm GeV}^2$.

Theoretically, pQCD (with OPE), in $\MSbar$ scheme, has been the usual approach to describe such quantities, cf.~\cite{EG1a,EG1b,DVCS}. This approach, however, has the theoretical disadvantage that the running coupling $a(Q^2)$ [$\equiv \alpha_s(Q^2)/\pi$] possesses Landau singularities at low positive $Q^2 \lesssim 0.1 \ {\rm GeV}^2$, and this makes it inconvenient for evaluation of spacelike observables ${\cal D}(Q^2)$ at low $Q^2$, such as BSR. In recent years, an extension of pQCD couplings to low $Q^2$, without Landau singularities, called (Fractional) Analytic Perturbation Theory [(F)APT)] \cite{ShS,MS96,ShS98,Sh,BMS05,BKS05,BMS06,BMS10,Bakulev} has been applied in the fitting of the theoretical OPE expression to the experimental inelastic contributions to BSR \cite{PSTSK10}, with good results.

In this work we fit the theoretical OPE expressions to the experimental BSR results in pQCD, in (F)APT, and two additional extensions of QCD to low $Q^2$, namely the 2$\delta$ \cite{2danQCD,mathprg2} and 3$\delta$ \cite{3l3danQCD,4l3danQCD} $\A$QCD. The latter two extensions have the coupling $\A(Q^2)$ [the analog of the pQCD coupling $a(Q^2)$] which is free of Landau singularities and physically motivated in the entire relevant regime of $Q^2$ in the complex plane, $Q^2 \in  \mathbb{C} \backslash (-\infty, -M_{\rm thr}^2]$, where $M_{\rm thr}^2 \lesssim 1 \ {\rm GeV}^2$ is a positive threshold scale. The present work is an extension of our previous work on BSR \cite{prev}; in comparison with the latter work, we now vary the $Q^2$-range of the fit, include the consideration of the elastic contribution, and estimate the uncertainties of the values of the extracted fit parameters due to systematic (in addition to statistical) uncertainties of the experimental BSR data.

The mentioned QCD variants (F)APT, 2$\delta$ and 3$\delta$ $\A$QCD were constructed directly by imposing certain physically motivated conditions on the QCD coupling. In this context, we point out that there exist several other approaches to the construction of the QCD coupling, among them those using Dyson-Schwinger equations which involve various versions of the dynamically generated gluon mass (the mentioned 2$\delta$ $\A$QCD gives a coupling with similar properties). For a recent review of various approaches, we refer to \cite{Brodrev}.
  
  In Sec.~\ref{sec:BSR} we present the theoretical leading-twist and higher-twist OPE contributions for the considered quantity BSR, as well as the parametrization of the elastic contribution to BSR. In Sec.~\ref{sec:num} we present the results of various fits of theoretical expressions to the experimental results. Finally, in Sec.~\ref{sec:concl} we summarize our conclusions.

Most of the formal aspects of the calculations are relegated to Appendices: in Appendix \ref{app:d1d2d3} we present the form of the leading-twist perturbation coefficients for a general renormalization scale and scheme; in Appendix \ref{app:An} we present construction of $\A_n$, the analogs of pQCD powers $a^n$, in extensions of pQCD without Landau singularities; in Appendix \ref{app:holA} we summarize such extensions [(F)APT, 2$\delta$ and 3$\delta$]; in Appendix \ref{app:fit} we explain how the statistical and systematic uncertainties of the experimental data are reflected in the corresponding uncertainties of the parameters extracted in the fits; and in Appendix \ref{app:Nf34} we estimate the effects of the finiteness of the charm quark mass in our evaluations. 

\section{Bjorken sum rule: theoretical expressions}
\label{sec:BSR}

The polarized Bjorken sum rule (BSR), $\Gamma_1^{p-n}$, is defined as the difference between proton and neutron polarized structure functions $g_1$ integrated over the whole $x$-Bjorken interval 
\be
\Gamma_1^{p-n}(Q^2)=\int_0^1 dx \left[g_1^p(x,Q^2)-g_1^n(x,Q^2) \right]\ .
\label{BSRdef}
\ee 
Based on the various measurements of these and the related structure functions, the inelatic part of the above quantity, $\Gamma_1^{p-n}(Q^2)_{\rm inel.}$, has been extracted at various values of squared momenta $Q_j^2$ ($0.054 \ {\rm GeV}^2 \leq Q_j^2 < 5 \ {\rm GeV}^2$): \cite{EG1a,EG1b,DVCS} (Jefferson Lab), \cite{E143,E155} (SLAC), \cite{COMP1,COMP2} (COMPASS at CERN). \footnote{The index $j$ in $Q_j^2$ indicates from here on that these are the scales at which the experimental values are given.}

Theoretically, this quantity can be written in the Operator Product Expansion (OPE) form \cite{BjorkenSR}
\be
\label{BSR}
\Gamma_1^{p-n, \rm {OPE}}(Q^2)= {\Big |}\frac{g_A}{g_V} {\Big |} \frac{1}{6}(1 - {\cal D}_{\rm BS}(Q^2))+\sum_{i=2}^\infty
\frac{\mu_{2i}(Q^2)}{Q^{2i-2}}\ .
\ee
Here, $|g_A/g_V|$ is the ratio of the nucleon axial charge, $(1-{\cal D}_{\rm BS})$ is the perturbation expansion for the leading-twist contribution, and $\mu_{2i}/Q^{2i-2}$ are the higher-twist contributions. The value obtained from neutron $\beta$ decay is known to a high accuracy, $|g_A/g_V|=1.2723 \pm 0.0023$ \cite{PDG2016}, and we will use the central value. In the higher-twist terms, we will take only  the terms $\sim 1/Q^2$ and $1/(Q^2)^2$.

\subsection{Perturbation expansion of the leading-twist}
\label{subs:PT}

The leading-twist term has the canonical part ${\cal D}_{\rm BS}(Q^2)$ whose perturbation expansion in $a \equiv \alpha_s/\pi$ is known up to N$^3$LO ($\sim a^4$)
\bea
{\cal D}_{\rm BS}(Q^2)_{\rm pt} & = & {\bar a} + {\bar d}_1 {\bar a}^2 + {\bar d}_2 {\bar a}^3 + {\bar d}_3 {\bar a}^4 + {\cal O}({\bar a}^5),
\label{DBSMS}
\eea
where the bar indicates that the expansion is in the $\MSbar$ scheme, and the renormalization scale $\mu^2$ is implicitly understood to be equal to the physical scale $Q^2$. The NLO, N$^2$LO and N$^3$LO coefficients ${\bar d}_j$ ($j=1,2,3$) were obtained in \cite{nloBSR,nnloBSR,nnnloBSR}, respectively. In the considered range of momentum transfer $0 < Q^2 < 5  \ {\rm GeV}^2$, we will assume that the effective number of active quark flavors is $N_f=3$, and therefore only the nonsinglet (NS) contributions appear.\footnote{The singlet contribution apeears for the first time at $\sim a^4$, and only if $N_f \not= 3$ \cite{Baikov:2015tea,Larin:2013yba}.}

When the renormalization scale is changed from $\mu^2=Q^2$ to a general value $\mu^2 = k Q^2$ ($0 < k \sim 1$), and when the renormalization scheme parameters are changed from the $\MSbar$ values ${\bar c}_j \equiv {\bar \beta}_j/\beta_0$ to general scheme parameter values $c_j$ ($j \geq 2$), the perturbation expansion changes accordingly 
\bea
{\cal D}_{\rm BS}(Q^2)_{\rm pt} & = & a(k Q^2) + d_1(k) a(k Q^2)^2 + d_2(k;c_2) a(k Q^2)^3 + d_3(k; c_2, c_3) a(k Q^2)^4 + {\cal O}(a^5).
\label{DBSpt}
\eea
The expressions for the new coefficients $d_1(k)$, $d_2(k; c_2)$ and $d_3(k; c_2, c_3)$ are obtained on the basis of independence of the observable ${\cal D}_{\rm BS}(Q^2)_{\rm pt}$ from $k$ and $c_j$ ($j \geq 2$), and are given in Appendix \ref{app:d1d2d3}.

In those versions of $\A$QCD where the coupling is a holomorphic function $\A(Q^2)$ [instead of the nonholomorphic $a(Q^2)$] with nonperturbative contributions, the power expansion (\ref{DBSpt}) becomes a nonpower expansion where $a^n$ get replaced by $\A_n$ ($\not= \A^n$)
\bea
{\cal D}_{\rm BS}(Q^2)_{\rm {\A}QCD} & = & \A(k Q^2) + d_1(k) \A_2(k Q^2) + d_2(k;c_2) \A_3(k Q^2) + d_3(k; c_2, c_3) \A_4(k Q^2) + {\cal O}(\A_5).
\label{DBSAQCD}
\eea
The construction of the power analogs $\A_n$ of $a^n$ were obtained in Ref.~\cite{CV1,CV2} for integer $n$ and in Ref.~\cite{GCAK} for general real $n > -1$. A brief description for the construction of $\A_n$ is given in Appendix \ref{app:An}. These expressions are based on the renormalization group equation (RGE), in close analogy with the RGE in the perturbation theory. The couplings $\A_n(Q^2)$ can be obtained once the coupling $\A(Q^2)$ is known. The construction of $\A(Q^2)$ coupling is summarized in Appendix \ref{app:holA} for various variants of QCD with holomorphic coupling: (F)APT, 2$\delta$ and 3$\delta$ $\A$QCD, and we refer to that Appendix for more details.

\subsection{Higher-twist}
\label{subs:HT}

In the theoretical OPE expression (\ref{BSR}), the term with the dimension $D=2$ (i.e., $\propto 1/Q^2$) has the coefficient
\be
\mu_4 = \frac{M_N^2}{9} (a_2^{p-n} + 4 d_2^{p-n} + 4 f_2^{p-n}(Q^2)),
\label{mu4}
\ee
where $M_N \approx 0.94$ GeV is the nucleon mass, $a_2^{p-n}$ is the (twist-2) target mass correction, and $d_2^{p-n}$  is a
twist-3 matrix element
\be
d_2^{p-n} = \int_0^1 dx x^2 (2 g_1^{p-n} + 3 g_2^{p-n}).
\label{d2pn}
\ee
At $Q^2=1 \ {\rm GeV}^2$, we have $a_2^{p-n}=0.031 \pm 0.010$ and $d_2^{p-n}=0.008 \pm 0.0036$ \cite{DVCS}. We will neglect $Q^2$-dependence of these two quantities [as was done also in Ref.~\cite{DVCS}], and will take the central values, i.e., $a_2^{p-n} + 4 d_2^{p-n}=0.063$. On the other hand, the coefficient $f_2^{p-n}(Q^2)$ will be a parameter of the fit, and its $Q^2$-dependence will not be neglected, its evolution is known \cite{ShuVa,KUY} in pQCD
\be
f_2^{p-n}(Q^2)=f_2^{p-n}(Q_{\rm in}^2) \left(\frac{a(Q^2)}
{a(Q_{\rm in}^2)}\right)^{\gamma_0/8\beta_0} \ ,
\label{HTQ2pt}
\ee
where $\nu \equiv \gamma_0/(8\beta_0) =32/81$ when $N_f=3$, and the reference scale will be taken $Q_{\rm in}^2 = 1 \ {\rm GeV}^2$. In QCD with holomorphic coupling $\A(Q^2)$, the power $a^{\gamma_0/8\beta_0}$ gets replaced by $\A_{\gamma_0/8\beta_0}$
(which is in general not equal to the simple power $\A^{\gamma_0/8\beta_0}$, as mentioned above, cf.~also Appendix \ref{app:An},
\be
f_2^{p-n}(Q^2)=f_2^{p-n}(Q_{\rm in}^2) \left(\frac{A_{\gamma_0/8\beta_0}(Q^2)}
{\A_{\gamma_0/8\beta_0}(Q_{\rm in}^2)}\right) \ ,
\label{HTQ2AQCD}
\ee

In addition, in some of the fits we will also include the $D=4$ term in the theoretical OPE expression (\ref{BSR}) $\mu_6/(Q^2)^2$, where we will consider the coefficient $\mu_6$ as $Q^2$-independent. Thus, our theoretical expression for BSR will be the truncated (at $D=4$, or $D=2$) OPE expression
\bea
\label{BSROPE}
\Gamma_1^{p-n,{\rm OPE[4]}}(Q^2; k, f_2^{p-n}(1); \mu_6) &=&
{\Big |} \frac{g_A}{g_V} {\Big |} \frac{1}{6}(1 - {\cal D}_{\rm BS}(Q^2)) +
\nonumber\\ &&
+ \frac{M_N^2}{Q^2} \frac{1}{9} \left( a_2^{p-n} + 4 d_2^{p-n} + 4 f_2^{p-n}(Q^2) \right) + \frac{\mu_6}{(Q^2)^2},
\eea
where the leading-twist truncated expressions are given in Eqs.~(\ref{DBSpt}) and (\ref{DBSAQCD}), and the $Q^2$-dependent part of the $D=2$ term (twist-4) in Eqs.~(\ref{HTQ2pt}) and (\ref{HTQ2AQCD}), for the pQCD and $\A$QCD version of QCD, respectively. We will regard the renormalization scale parameter $k \equiv \mu^2/Q^2$ and the higher-twist coefficients $f_2^{p-n}(Q^2_{\rm in})$ (with $Q^2_{\rm in}=1 \ {\rm GeV}^2$) and $\mu_6$ as the free parameters to be determined in the fitting procedure.

In our approach, we will include in the leading-twist part of the OPE all the known terms (i.e., up to order $\sim a^4 \sim \A_4$). The order of the leading-twist terms in general affects the fitted higher-twist contributions (cf.~\cite{Kataev:1997nc,Narison:2009ag,Parente:1994bf} for the fit of truncated OPE to structure functions).

\subsection{Elastic contribution}
\label{subs:EC}

The OPE evaluation is, in principle, for inclusive observables; in the case of BSR, this means that the OPE fit should be applied to the experimental values of the sum of the inelastic and elastic contribution. 

The elastic contribution to BSR can be expressed \cite{Osip,Merg} in terms of the proton and neutron electromagnetic form factors $F_1$ and $F_2$ 
\be
\Gamma_1^{p-n}(Q^2)_{\rm el.}=\frac{1}{2} F_1^{p}(Q^2) \left( F_1^{p}(Q^2) + F_2^{p}(Q^2) \right) -  \frac{1}{2} F_1^{n}(Q^2) \left( F_1^{n}(Q^2) + F_2^{n}(Q^2) \right).\label{G1el1} 
\ee
The most recent parametrization of these form factors was performed in \cite{Sabbiretal} from light-front holographic QCD (LFH QCD).
Namely, these form factors can be expressed in terms of the inverse power expressions
\be
{\cal F}_{\tau} \equiv \left( 1 + \frac{Q^2}{M_{\rho}^2} \right)^{-1} \times ... \times \left( 1 + \frac{Q^2}{(2 \tau -3 ) M_{\rho}^2} \right)^{-1}, \qquad (\tau=2,3,\ldots),
\label{Ftau}
\ee
which are products of $\tau-1$ poles along the vector meson Regge radial trajectory in terms of the $\rho$-vector mass $M_{\rho}=0.755$ GeV and its radial excitations. We have
\bes
\label{Fj}
\bea
F_1^p(Q^2) &=& {\cal F}_{\tau=3}(Q^2), \qquad
F_2^p(Q^2) = \chi_p \left[ (1 - \gamma_p) {\cal F}_{\tau=4}(Q^2) + \gamma_p{\cal  F}_{\tau=6}(Q^2) \right],
\label{Fjp}
\\
F_1^n(Q^2) &=& -\frac{1}{3} r \left[ {\cal F}_{\tau=3}(Q^2) - {\cal F}_{\tau=4}(Q^2) \right],
  \qquad
F_2^n(Q^2) = \chi_n \left[ (1 - \gamma_n) {\cal F}_{\tau=4}(Q^2) + \gamma_n{\cal  F}_{\tau=6}(Q^2) \right].
\label{Fjn}
\eea
\ees
Here: $\chi_p = \mu_p-1 = 1.793$ and $\chi_n = \mu_n=-1.913$ are the anomalous moments of $p$ and $n$, respectively; $\gamma_{n,p}$ are the higher Fock probabilities for the $L=0 \to L=1$ spin-flip electromagnetic form factors, and $r$ is a phenomenological factor. The values of these three parameters are obtained by fitting to polarization data for the form factors \cite{Sabbiretal}: $\gamma_p=0.27$; $\gamma_n=0.38$; $r=2.08$.

As a consequence, the elastic contribution (\ref{G1el1}) can be represented as a combination of inverse powers  $( 1 + Q^2/M_{\rho}^2/(2 m -1) )$, and for large $Q^2$ this can be expanded in inverse powers of $Q^2$
\be
\Gamma_1^{p-n}(Q^2)_{\rm el.} = 2.3368 \left(\frac{M_{\rho}^2}{Q^2}\right)^4 + 13.8851 \left(\frac{M_{\rho}^2}{Q^2}\right)^5 + {\cal O}(1/Q^{12}).
\label{G1elexp}
\ee

This means that at high $Q^2$ the elastic contribution starts with the dimension $D=8$ term $\sim (1/Q^2)^4$. Theoretically, it is not included in the truncated OPE expression (\ref{BSROPE}) in pQCD.\footnote{However, the last coefficient in the truncated OPE, e.g.~$\mu_6$, is sometimes in the literature regarded to include, in a certain effective way, the contributions from the higher dimensional terms $D \geq 6$, cf.~\cite{DVCS} for a discussion of these aspects.}$^{,}$\footnote{The elastic contributions parametrized in a way different from that of Ref.~\cite{Sabbiretal} could in general contain terms of dimension $D<8$. Nonetheless, the fitted expressions of Ref.~\cite{Merg} give for the elastic contributions an expansion similar to Eq.~(\ref{G1elexp}), where the first nonzero term is a very small $D=6$ contribution which is negligible in comparison with $D=8$ term (the coefficient at $D=6$ term is by about a factor of $10^{-4}$ smaller than at $D=8$ term). All these terms there are divided by a large weakly $Q^2$-dependent factor $[\ln(27.81 + 1.72 \; Q^2/M_{\rho}^2)]^{4.296}$.}
Further, in 3$\delta$ and 2$\delta$ $\A$QCD, the higher dimensional terms up to (and including) the $D=8$ term are not included in the leading-twist contribution (\ref{DBSAQCD}) because in these approaches $\A(Q^2) - a(Q^2) \sim (\Lambda^2/Q^2)^5$ as explained in Appendix \ref{subsapp:2d3d} Eq.~(\ref{diffAaN5}). Therefore, the truncated  OPE expression (\ref{BSROPE}) does not contain the dimension $D=8$ term $\sim (1/Q^2)^4$ also in 3$\delta$ and 2$\delta$ $\A$QCD approaches. The only exception is the (F)APT where $\A(Q^2) - a(Q^2) \sim (\Lambda^2/Q^2)^1$, cf.~Appendix \ref{subsapp:FAPT}, and where the truncated OPE (\ref{BSROPE}) could contain, in principle, all the elastic contributions, including the $\sim (1/Q^2)^4$ term.

Therefore, it looks more natural to first fit the truncated OPE expression (\ref{BSROPE}), with pQCD and  (3$\delta$ and 2$\delta$) $\A$QCD, to the BSR experimental results \cite{EG1a,EG1b,DVCS,E143,COMP2} which are obtained for the inelastic contribution; and after such a fit, add the elastic contribution (\ref{G1el1}) [with the parametrization (\ref{Ftau})-(\ref{Fjn})]. We note that the theoretical expression obtained for the total BSR in this way is the sum of the expressions (\ref{BSROPE}) and (\ref{G1el1}), which is again an OPE expression as it should be for such an inclusive spacelike observable as the total BSR. 

The other approach would be to fit the truncated OPE expression (\ref{BSROPE}) with the experimental points for the total BSR, i.e., fit with the experimental points for inelastic BSR with the elastic contribution added to them. Such an approach seems less natural, at least in pQCD and  3$\delta$ and 2$\delta$ $\A$QCD, because the truncated expression (\ref{BSROPE}) in these cases in principle does not contain the leading elastic contribution $\sim 1/(Q^2)^4$, cf.~Eq.~(\ref{diffAaN5}) in Appendix \ref{subsapp:2d3d}.

Nonetheless, below we will apply, for completeness, both approaches in our numerical fitting procedures.

\section{Numerical fits}
\label{sec:num}

In the numerical fits, we will consider the following parameters to be fitted in the truncated OPE expression (\ref{BSROPE}): (i) the renormalization scale parameter $k \equiv \mu^2/Q^2$ of the (truncated) leading-twist contribution (\ref{DBSpt}) [for pQCD] or (\ref{DBSAQCD}) [for $\A$QCD]; (ii) the parameter $f_2^{p-n}(1 \ {\rm GeV}^2)$ appearing in the $D=2$ term of Eq.~(\ref{BSROPE}) [cf.~Eqs.~(\ref{HTQ2pt}) and (\ref{HTQ2AQCD})]; (iii) and in some fits the parameter $\mu_6$ in the $D=4$ term of Eq.~(\ref{BSROPE}) will be included. We will take for the experimental data points for the inelastic contribution to BSR the data from \cite{EG1a,EG1b,DVCS,E143,COMP2};\footnote{
  We will not take into account the SLAC E155, SMC and HERMES points \cite{E155,SMC,HERMES}, at $Q^2=5 \ {\rm GeV}^2$, $10 \ {\rm GeV}^2$, and $2.5$ (and $5$) ${\rm GeV}^2$ , respectively, as they were obtained by NLO pQCD $Q^2$-evolution from data distributed across a wide range of $Q^2$. Among the points of SLAC E143 we exclude the point at $Q^2=5 \ {\rm GeV}^2$ for the same reason.}
they are in the momentum interval $0.054 \ {\rm GeV}^2 \leq Q_j^2 < 5 \ {\rm GeV}^2$. Among these data points, we exclude from the fit the four points with $Q_j^2 \geq 3 \ {\rm GeV}^2$ (three from \cite{DVCS}, at $Q_j^2=3.223$, $3.871$ and $4.739 \ {\rm GeV}^2$; and one from \cite{COMP2} at $Q^2=3 \ {\rm GeV}^2$) because they tend to decrease the quality parameter $\chi^2/{\rm (d.o.f.)}$ of the fit significantly. Nonetheless, we will show both quality parameters  $\chi^2/{\rm (d.o.f.)}$ obtained from the mentioned fits, i.e., the one with the four points excluded, and included.\footnote{If we perform the fit with the four points included, the results differ somewhat, but not significantly.} The number of active flavors will be kept all the time at $N_f=3$.
The fits will be performed by the least squares method, taking into account the statistical uncertainties $\sigma_{j,{\rm stat}}$ in the points $Q_j^2$, we will consider them independent of each other. These uncertainties $\sigma_{j,{\rm stat}}$ are in general significantly smaller than the systematic uncertainties $\sigma_{j,{\rm sys}}$. The latter are strongly correlated, and we will consider them as completely correlated. The uncertainties in the values of the extracted parameters $k \equiv \mu^2/Q^2$, $f_2^{p-n}(Q^2_{\rm in})$ and $\mu_6$ are then due to statistical (small) and systematic (larger) uncertainties of the data. We refer to Appendix \ref{app:fit} on how we obtained the uncertainties of the extracted values of the fit parameters.

Using specific values of the renormalization scale parameter $k \equiv \mu^2/Q^2$ may allow us to incorporate in our evaluations (\ref{DBSAQCD}) at least a part of the contributions of higher orders of the series. It is expected that $k \sim 1$, and usually it is taken in the literature in the range $1/2 < k < 2$, sometimes $1/4 < k < 4$. In the considered work, after replacing the powers of the pQCD coupling by their analytic counterparts, cf.~Eqs.~(\ref{DBSpt})-(\ref{DBSAQCD}), a spacelike observable depends usually weakly on the contributions of higher orders. Still, in order not to miss the possibly relevant influence of higher orders, we decided to increase the range of possible $k$ values to: $1/16 < k < 16$. This choice still avoids very large or very small renormalization scales where the corresponding coefficients $d_n(k)$ at the powers $a(k Q^2)^{n+1}$, or at their analytic counterparts $\A_{n+1}(k Q^2)$, contain powers of the large terms $\sim \ln k$ [cf.~Eqs. (\ref{d1d2d3})] which may destroy the convergence of the series already at low $n$. We will see that the results of the fitting will rarely give the extreme values $k=16$ or $k=1/16 \approx 0.063$, mostly in the cases of ($\MSbar$) pQCD and (F)APT.

We present below the results of the fits of the inelastic contributions to the truncated OPE expression (\ref{BSROPE}), first in Sec.~\ref{subs:mu60} with $\mu_6=0$, and then in Sec.~\ref{subs:mu6} for $\mu_6 \not=0$ included as a fit parameter; the elastic contribution is added afterwards. In Sec.~\ref{subs:mass} we present the corresponding fits for the case when the higher-twist term in the OPE has a mass parameter. In Sec.~\ref{subs:lowQ2} we include the fits for two different ans\"atze for BSR at very low $Q^2$. Finally, in Sec.~\ref{subs:tot} we present the fits of the truncated OPE expression (\ref{BSROPE}), with $\mu_6 \not=0$ included as a fit parameter, applied to the total BSR, i.e., to the sum of the inelastic data points (at $Q_j^2$) and the corresponding elastic contribution $\Gamma_1^{p-n}(Q_j^2)_{\rm el.}$.

In each case, the fits are performed for four variants of QCD: in pQCD (in $\MSbar$), in (F)APT (in $\MSbar$), and in 2$\delta$ \cite{mathprg2} and 3$\delta$ $\A$QCD \cite{4l3danQCD}. Each of these fits is performed by excluding among the data points those with $Q^2 < Q^2_{\rm min}$, where $Q^2_{\rm min}=0.268$ or $0.66 \ {\rm GeV}^2$ (and sometimes also:  $Q^2_{\rm min}=0.47 \ {\rm GeV}^2$). As mentioned earlier, the four points at high $Q^2 \geq 3 \ {\rm GeV}^2$ are excluded from the fit as well.

It is reasonable to assume that the number of active quark flavors in the considered interval $0.054 \ {\rm GeV}^2 < Q^2 < 5 \ {\rm GeV}^2$ is $N_f=3$, cf.~Appendix \ref{app:Nf34}. All the fits will be performed in $\MSbar$ pQCD approach, in the Fractional Analytic Perturbation Theory [(F)APT], and in 2$\delta$ and 3$\delta$ $\A$QCD approach, cf.~Appendix \ref{app:holA} where also the values of the parameters of these three QCD variants are presented. In $\MSbar$ pQCD and in 2$\delta$ and 3$\delta$ $\A$QCD the (underlying) pQCD running coupling $a(Q^2)$ is determined by the requirement $\pi a(M_Z^2;\MSbar) = 0.1185$ \cite{PDG2014,PDG2016}, and in (F)APT we use ${\overline \Lambda}_3=0.45$ GeV (at the end of Sec.~\ref{subs:mu6} we also comment on the case ${\overline \Lambda}_3=0.40$ GeV); we refer to Appendix \ref{app:holA} for more details.

\subsection{Fits with $\mu_6=0$}
\label{subs:mu60}

\begin{figure}[htb] 
\begin{minipage}[t]{0.49\linewidth}
\centering\includegraphics[width=89mm]{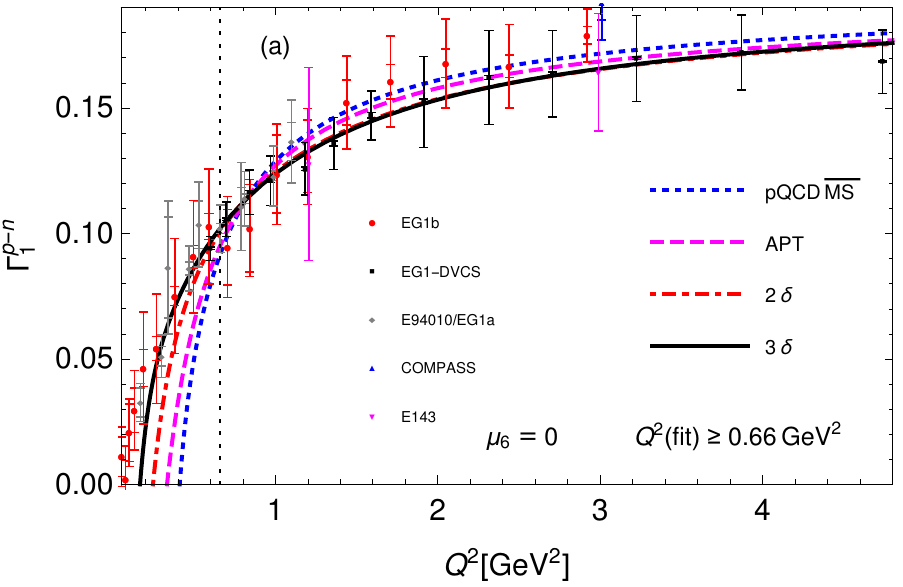}
\end{minipage}
\begin{minipage}[t]{0.49\linewidth}
\centering\includegraphics[width=89mm]{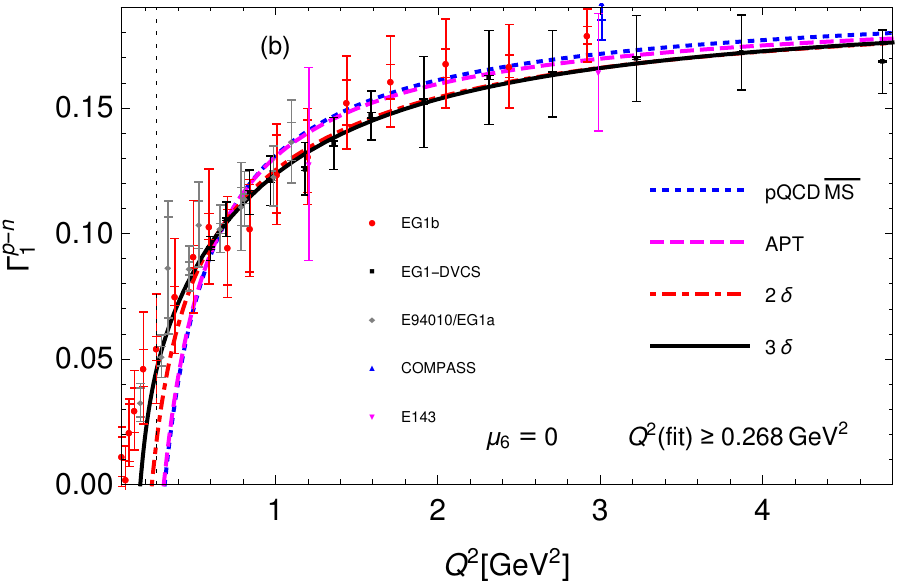}
\end{minipage}
\vspace{-0.4cm}
\caption{(color online): Fits of the OPE expression (\ref{BSROPE}) truncated at $D=2$ term, i.e., $\mu_6=0$, to the experimental data for $\Gamma_1^{p-n}(Q^2)_{\rm inel.}$, in four different QCD variants, where in the fits: (a) $Q^2 \geq 0.66 \ {\rm GeV}^2$; (b) $Q^2 \geq 0.268 \ {\rm GeV}^2$. See the text for details. The respective lower bound of the fitting interval, $Q^2_{\rm min}=0.66$ or $0.268 \ {\rm GeV}^2$, is included as the thin dotted vertical line.}
\label{Figmu60}
\end{figure}

In Figs.~\ref{Figmu60}(a),(b), we present the curves in the mentioned four QCD variants, for $Q^2_{\rm min}=0.66$ and $0.268 \ {\rm GeV}^2$, respectively.
The corresponding results and four fit quality parameters $\chi^2/{\rm d.o.f}$ are given in Table \ref{tabmu60}, for $Q^2_{\rm min} = 0.66$, $0.47$ and $0.268 \ {\rm GeV}^2$ .
\begin{table}
\caption{The values of the extracted fit parameters $k = \mu^2/Q^2$ and $f_2^{p-n}(Q^2_{\rm in})$ ($Q^2_{\rm in} = 1 \ {\rm GeV}^2$), obtained by fitting the OPE expression (\ref{BSROPE}), truncated at $D=2$ ($\mu_6=0$), for the considered four QCD variants, and for the minimal $Q^2$ values of the fit being $0.66$, $0.47$ and $0.268 \ {\rm GeV}^2$. The uncertainties of the extracted $f_2^{p-n}$ values are separated to those coming from the statistical (small) and systematic (large) uncertainties of the experimental data (cf.~Appendix \ref{app:fit}).
The resulting fit quality parameters $\chi^2/{\rm d.o.f.}$ are given as well (see the text for explanation).} 
\label{tabmu60}
\begin{ruledtabular}
\centering
\begin{tabular}{r l|llllll}
  QCD variant & $Q^2_{\rm min}({\rm fit})$ & $k$ & $f_2^{p-n}(1.)$ & $\chi^2/{\rm d.o.f}$ & $\chi^2_{\rm ext}/{\rm d.o.f}$ & $\chi^2_{0.268}/{\rm d.o.f.}$ & $\chi^2_{\rm all}/{\rm d.o.f.}$
\\
\hline
$\MSbar$ pQCD & 0.66 & 6.04 & $-0.103 \pm 0.001 \pm 0.011$ & 44.5 & 161. & $6.81 \times 10^4$ & $\infty$
\\
(F)APT       & 0.66 & 2.74 & $-0.143 \pm 0.001 \pm 0.021$ & 23.3 & 85.6 & 63.0 & 970.
\\
2$\delta$    & 0.66 & 0.397 & $-0.077 \pm 0.001 \pm 0.024$ & 8.34 & 52.2 & 11.1 & 406.
\\
3$\delta$ & 0.66 & 2.41 & $-0.064 \pm 0.001 \pm 0.023$ & 6.85 & 56.1 & 6.17 & 130.
\\
\hline
$\MSbar$ pQCD & 0.47 & 6.35 & $-0.094 \pm 0.000 \pm 0.009$ & 66.1 & 177. & $5.71 \times 10^4$ & $\infty$
\\
(F)APT       & 0.47 & 3.03 & $-0.137 \pm 0.000 \pm 0.015$ & 30.5 & 88.4 & 51.2 & 873.
\\
2$\delta$    & 0.47 & 0.395 & $-0.077 \pm 0.000 \pm 0.023$ & 7.91 & 45.2 & 10.9 & 402.
\\
3$\delta$ & 0.47 & 2.58 & $-0.064 \pm 0.000 \pm 0.023$ & 6.49 & 50.2 & 5.92 & 129.
\\
\hline
$\MSbar$ pQCD & 0.268 & 16. & $-0.160 \pm 0.001 \pm 0.045$ & 58.7 & 145. & 58.7 & $1.20 \times 10^{3}$
\\
(F)APT       & 0.268 & 3.71 & $-0.134 \pm 0.000 \pm 0.102$ & 49.6 & 103. & 49.6 & 827.
\\
2$\delta$    & 0.268 & 0.338 & $-0.074 \pm 0.000 \pm 0.022$ & 10.6 & 46.6 & 10.6 & 372.
\\
3$\delta$ &  0.268 & 2.55 & $-0.064 \pm 0.000 \pm 0.018$ & 5.91 & 44.5 & 5.91 & 129.
\end{tabular}
\end{ruledtabular}
\end{table}
As mentioned earlier, the fit is performed for the data in the momentum interval $Q^2_{\rm min} \leq Q_j^2 \leq 3 \ {\rm GeV}^2$ (with the COMPASS data point \cite{COMP2} excluded), and the corresponding quality parameter for that interval is denoted as $\chi^2/{\rm d.o.f}$. In addition, $\chi^2_{\rm ext}/{\rm d.o.f.}$ is the quality parameter for the wider interval $Q^2_{\rm min} \leq Q_j^2 < 5 \ {\rm GeV}^2$; $\chi^2_{0.268}/{\rm d.o.f.}$ is the parameter when  $0.268 \ {\rm GeV}^2 \leq Q_j^2 \leq 3 \ {\rm GeV}^2$; and $\chi^2_{\rm all}/{\rm d.o.f.}$ is the parameter when  all the experimental points \cite{EG1a,EG1b,DVCS,E143,COMP2} are included, $0.054 \ {\rm GeV}^2 \leq Q_j^2 < 5 \ {\rm GeV}^2$. In these quantities $\chi^2$, the central values of $k$ and $f_2^{p-n}(Q^2_{\rm in})$ ($Q^2_{\rm in} = 1 \ {\rm GeV}^2$) parameters obtained from the mentioned fit interval $Q^2_{\rm min} \leq Q_j^2 \leq 3 \ {\rm GeV}^2$ are used; we take here ``d.o.f'' as $(N-p)$ where $N$ is the number of fitted data points entering the considered $\chi^2$ and $p$ is the number of parameters of the fit ($p=2$ here)
\be
\chi^2(Q_{K+1}^2 \leq Q^2 \leq Q_{K+N}^2)/{\rm d.o.f.} = \frac{1}{(N-p)}\sum_{j=K+1}^{K+N} \frac{1}{\sigma^2_{j,{\rm stat}}} \left(
\Gamma_1^{p-n,OPE[4]}(Q_j^2; k, f_2^{p-n}(1); \mu_6=0) -  \Gamma_1^{p-n}(Q_j^2)_{\rm exp} \right)^2.
\label{chi2def}
\ee
The values of these quantities are always large, in the best cases between 1 and 10. This is so because the statistical uncertainties of the newer JLAB data \cite{DVCS} are very small, $\sigma_{j,{\rm stat}} \lesssim 10^{-3}$, and simultaneously, our fit function (truncated OPE) is different from the ideal function which we do not know. Nonetheless, they decrease when analytic QCD variants are employed, especially 2$\delta$ and 3$\delta$ $\A$QCD.

We wish to point out that the approach of ($\MSbar$) QCD in the case of $Q_{\rm min}^2=0.268 \ {\rm GeV}^2$ is, in principle, not applicable. This is so because the corresponding coupling $a(Q^2)$ has a Landau branching point at $Q^2_{\rm branch}=0.371 \ {\rm GeV}^2$, which makes the $D=2$ running coefficient $f_2^{p-n}(Q^2)$, Eq.~(\ref{HTQ2pt}), undefined at $Q^2 \leq 0.371 \ {\rm GeV}^2$. Nonetheless, in order to be able to present a curve, we applied in the fitting case $Q_{\rm min}^2=0.268 \ {\rm GeV}^2$ in the $\MSbar$ pQCD approach a restriction on the leading-twist renormalization scale $\mu^2 = k Q^2$, namely $k > 1.383$; and this not just in the leading-twist contribution, but we also made an ad hoc replacement in the $D=2$ running coefficient $f_2^{p-n}(Q^2)$, Eq.~(\ref{HTQ2pt}): $f_2^{p-n}(Q^2) \mapsto f_2^{p-n}(k Q^2)$. We applied this also in the case when $\mu_6 \not= 0$ (Sec.~\ref{subs:mu6}, for  $\MSbar$ pQCD approach with  $Q_{\rm min}^2=0.268 \ {\rm GeV}^2$). In other approaches (APT and $\A$QCD's) this is not necessary, as there are no Landau singularities. Further, we note that in the last column in Table \ref{tabmu60} we have $\chi^2/{\rm d.o.f.} = \infty$ in the case of $\MSbar$ pQCD for $Q^2_{\rm min}=0.66$ and $0.47 \ {\rm GeV}^2$. This is so because at the lowest available experimental point, $Q_{j=1}^2=0.054 \ {\rm GeV}^2$, the coupling is $a(k Q_1^2; \MSbar) = \infty$ because $k Q_1^2 < Q^2_{\rm branch}=0.371 \ {\rm GeV}^2$ and thus we hit Landau singularities there.

We can deduce from Figs.~\ref{Figmu60} and Table \ref{tabmu60}: (a) the best results in the considered approach ($\mu_6=0$) are obtained in 3$\delta$ $\A$QCD; (b) the quality of extrapolation of the obtained fitted curves from the fitting interval, $Q^2_{\rm min} \leq Q^2 \leq 3 \ {\rm GeV}^2$, to the entire interval, $0.054 \ {\rm GeV}^2 \leq Q^2 < 5 \ {\rm GeV}^2$, does not improve significantly when the fitted interval is extended (i.e., when $Q^2_{\rm min}$ is lowered), cf.~also the last column in Table \ref{tabmu60}. \textcolor{black}{This indicates that the behavior of the curves in the extrapolated regions, $0.054 \ {\rm GeV}^2 < Q^2 < 0.268 \ {\rm GeV}^2$ and $3 \ {\rm GeV}^2 < Q^2 \leq 5 \ {\rm GeV}^2$, is of similar quality in the cases of different values of $Q^2_{\rm min}({\rm fit})$, and gives the dominant part of $\chi^2_{\rm all}/{\rm d.o.f.}$. The same can be observed for the values of $f_2^{p-n}(1)$: they do not depend much on the value of  $Q^2_{\rm min}({\rm fit})$, but only on the QCD variant used in the fit. This appears to be related with the fact that only one parameter beyond the leading-twist contribution is used here ($\mu_4 \leftrightarrow f_2^{p-n}$), representing a truncated OPE ansatz with a significantly restricted freedom.}

\subsection{Fits with $\mu_6 \not= 0$}
\label{subs:mu6}

\begin{figure}[htb] 
  \begin{minipage}[t]{0.49\linewidth}
\centering\includegraphics[width=89mm]{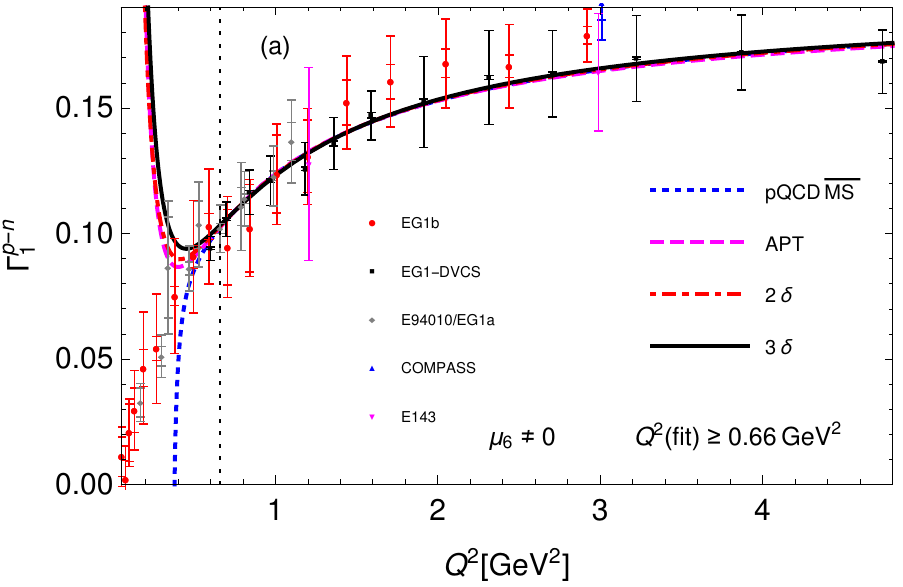}
\end{minipage}
\begin{minipage}[t]{0.49\linewidth}
\centering\includegraphics[width=89mm]{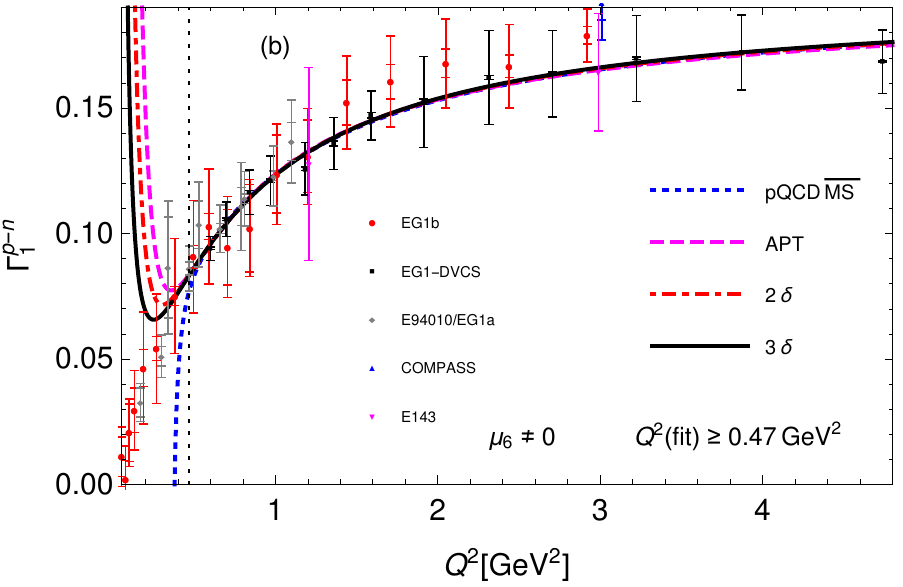}
\end{minipage}
\vspace{-0.4cm}
\caption{(color online): Fits of the OPE expression (\ref{BSROPE}) truncated at $D=4$ term (i.e., $\mu_4 \not=0$), to the experimental data for $\Gamma_1^{p-n}(Q^2)_{\rm inel.}$, done in four different QCD variants, where in the fits: (a) left plot, $Q^2 \geq 0.66 \ {\rm GeV}^2$; (b) right plot, $Q^2 \geq 0.47 \ {\rm GeV}^2$. See the text for details. The respective lower bound of the fitting interval, $Q^2_{\rm min}=0.66$ and $0.47 \ {\rm GeV}^2$, is included as the thin dotted vertical line.}
\label{Fig066047}
\end{figure}

\begin{figure}[htb] 
\begin{minipage}[t]{0.49\linewidth}
\centering\includegraphics[width=89mm]{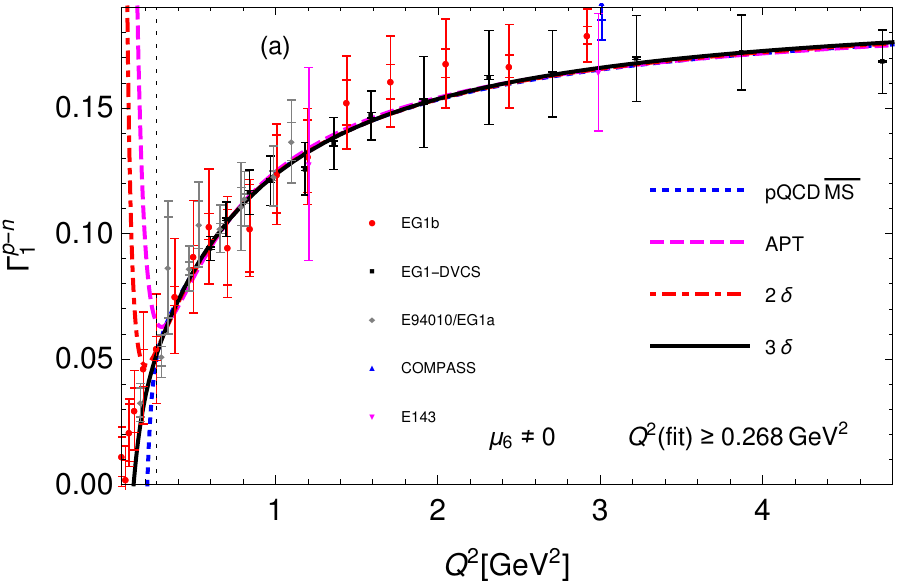}
\end{minipage}
\begin{minipage}[t]{0.49\linewidth}
\centering\includegraphics[width=89mm]{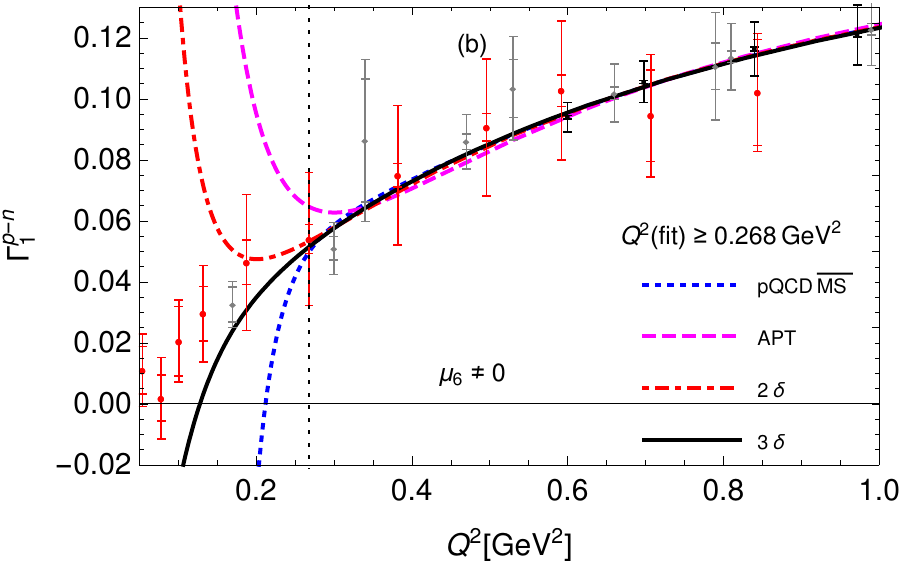}
\end{minipage}
\vspace{-0.4cm}
\caption{(color online): As Fig.~\ref{Fig066047}, but for  $Q^2_{\rm min} = 0.268 \ {\rm GeV}^2$: (a) for larger $Q^2$; (b) for $Q^2 < 1 \ {\rm GeV}^2$. The lower bound of the fitting interval, $Q^2_{\rm min}=0.268 \ {\rm GeV}^2$, is included as the thin dotted vertical line.}
\label{Fig0268}
\end{figure}

When we include $\mu_6$ in the fit as the third parameter, the resulting curves are presented in Figs.~\ref{Fig066047}(a),(b), for $Q^2_{\rm min}=0.66$ and $0.47 \ {\rm GeV}^2$, respectively. Further, when $Q^2_{\rm min}=  0.268 \ {\rm GeV}^2$, the results are shown in Figs.~\ref{Fig0268}(a),(b), at the higher $Q^2$ and the lower $Q^2 < 1 \ {\rm GeV}^2$ momenta, respectively. The corresponding results are given in Table \ref{tabmu6not0}, for $Q^2_{\rm min} = 0.66$, $0.47$ and $0.268 \ {\rm GeV}^2$, with the same notations as in Table \ref{tabmu60}.
\begin{table}
\caption{The values of the extracted fit parameters $k = \mu^2/Q^2$, $f_2^{p-n}(1 \ {\rm GeV}^2)$  and $\mu_6$ (in ${\rm GeV}^4$), obtained by fitting the OPE expression (\ref{BSROPE}), truncated at $D=4$ ($\mu_6 \not= 0$). The notations are as in Table \ref{tabmu60}. See the text in Sec.~\ref{subs:mu60} for explanation of the various $\chi^2/{\rm d.o.f.}$'s.} 
\label{tabmu6not0}
\begin{ruledtabular}
\centering
\begin{tabular}{r l|lllllll}
  QCD variant & $Q^2_{\rm min}({\rm fit})$ & $k$ & $f_2^{p-n}(1.)$ & $\mu_6$ & $\chi^2/{\rm d.o.f}$ & $\chi^2_{\rm ext}/{\rm d.o.f}$ & $\chi^2_{0.268}/{\rm d.o.f.}$ & $\chi^2_{\rm all}/{\rm d.o.f.}$
\\
\hline
$\MSbar$ pQCD & 0.66 & 16. & $-0.219 \pm 0.002 \pm 0.111$ & $0.032 \pm 0.001 \pm 0.055$ & 7.31 & 53.4 & $2.92 \times 10^5$ & $1.68 \times 10^{6}$
\\
(F)APT       & 0.66 & 0.063 & $-0.198 \pm 0.002 \pm 0.106$ & $0.018 \pm 0.001 \pm 0.028$ & 7.85 & 44.2 & 17.7 & $1.04 \times 10^{4}$
\\
2$\delta$    & 0.66 & 0.999 & $-0.116 \pm 0.002 \pm 0.094$ & $0.014 \pm 0.001 \pm 0.023$ & 6.99 & 51.8 & 19.5 & $5.92 \times 10^{3}$
\\
3$\delta$ & 0.66 & 4.46 & $-0.101 \pm 0.002 \pm 0.114$ &  $0.013 \pm 0.001 \pm 0.025$ &6.45 & 57.5 & 28.0 & $7.27 \times 10^{3}$
\\
\hline
$\MSbar$ pQCD & 0.47 & 16. & $-0.216 \pm 0.002 \pm 0.120$ & $0.031 \pm 0.001 \pm 0.045$ & 7.34 & 47.3 & $2.85 \times 10^5$ & $1.66 \times 10^{6}$
\\
(F)APT       & 0.47 & 0.063 & $-0.194 \pm 0.002 \pm 0.098$ & $0.016 \pm 0.001 \pm 0.021$ & 7.65 & 35.6 & 10.0 & $6.95 \times 10^{3}$
\\
2$\delta$    & 0.47 & 0.772 & $-0.098 \pm 0.002 \pm 0.078$ & $0.008 \pm 0.001 \pm 0.017$ & 6.88 & 48.2 & 7.42 & $1.40 \times 10^{3}$
\\
3$\delta$ & 0.47 & 3.10 & $-0.072 \pm 0.002 \pm 0.138$ &  $0.003 \pm 0.001 \pm 0.026$ &6.64 & 53.5 & 6.56 & 218.
\\
\hline
$\MSbar$ pQCD & 0.268 & 3.00 & $-0.180 \pm 0.001 \pm 0.237$ & $0.023 \pm 0.000 \pm 0.013$ & 6.52 & 38.7 & 6.52 & $\infty$
\\
(F)APT       & 0.268 & 5.27 & $-0.181 \pm 0.001 \pm 0.194$ & $0.013 \pm 0.000 \pm 0.011$ & 7.77 & 37.0 & 7.77 & $3.91 \times 10^{3}$
\\
2$\delta$    & 0.268 & 0.149 & $-0.089 \pm 0.001 \pm 0.166$ & $0.005 \pm 0.000 \pm 0.010$ & 6.27 & 41.9 & 6.27 & 293.
\\
3$\delta$ & 0.268 & 2.72 & $-0.007 \pm 0.001 \pm 0.262$ &  $0.001 \pm 0.000 \pm 0.015$ & 5.99 & 46.5 & 5.99 & 47.3
\end{tabular}
\end{ruledtabular}
\end{table}
The various versions of $\chi^2/{\rm d.o.f.}$ are those as explained in the previous Sec.~\ref{subs:mu60}, except that now in the relation (\ref{chi2def}) the factor in front of the sum is $1/(N-3)$ ($p=3$, d.o.f. is $N-3$).

Comparing Table \ref{tabmu6not0} and Figs.~\ref{Fig066047}-\ref{Fig0268} with Table \ref{tabmu60} and the Figs.~\ref{Figmu60} of the previous Sec.~\ref{subs:mu60} where $\mu_6=0$ was kept, we can see that in the cases of $Q^2({\rm fit}) \geq 0.66 \ {\rm GeV}^2$ and $Q^2({\rm fit}) \geq 0.47 \ {\rm GeV}^2$ the $\mu_6=0$ fits give in general better $\chi^2_{\rm all}/{\rm d.o.f.}$ (the last columns of Tables \ref{tabmu60} and \ref{tabmu6not0}). This means that the extrapolation down to the lowest experimental point $Q^2=Q_{j=1}^2=0.054 \ {\rm GeV^2}$ is better when $\mu_6=0$ (with the exception of $\MSbar$ pQCD case where problems with Landau singularities appear). This occurs because the inclusion of the $\mu_6/(Q^2)^2$ term in the truncated OPE makes this expression less stable at very low $Q^2$.

On the other hand, when  $Q^2({\rm fit}) \geq 0.268 \ {\rm GeV}^2$, some of the fits (2$\delta$ and 3$\delta$ $\A$QCD) give better extrapolation when $\mu_6 \not=0$. Table \ref{tabmu6not0} and Figs.~\ref{Fig0268} also show that when $\mu_6 \not= 0$ and $Q^2({\rm fit}) \geq 0.268 \ {\rm GeV}^2$, the best extrapolation to low $Q^2$ is obtained in 3$\delta$ $\A$QCD, followed by 2$\delta$ $\A$QCD. As Fig.~\ref{Fig0268}(b) suggests, the fitted curve in pQCD $\MSbar$ extrapolated to low $Q^2$ appears to be almost as good; in this case, however, we should keep in mind that the renormalization scale is $\mu^2  = k Q^2$ ($k =3.00$) and that this scale was used also in $f^{p-n}_2(Q^2)$, i.e., the ad hoc replacement $f^{p-n}_2(Q^2) \mapsto f^{p-n}_2(k Q^2)$ was performed in order to avoid the Landau singularities in the $D=2$ term at $Q^2 \geq 0.268 \ {\rm GeV}^2$ (cf.~also the discussion about that point in Sec.~\ref{subs:mu60}). Despite this replacement, in $\chi^2_{\rm all}/{\rm d.o.f.}$ Landau singularities are hit, because the three lowest experimental points give $k Q_{j}^2 < Q^2_{\rm branch}=0.371 \ {\rm GeV}^2$ ($j=1,2,3$) and are thus in the Landau singularity region.\footnote{We have $Q_1^2=0.054 \ {\rm GeV}^2$, $Q_2^2=0.078 \ {\rm GeV}^2$, and $Q_3^2=0.101 \ {\rm GeV}^2$.}

In the case of (F)APT, in contrast to 2$\delta$ and 3$\delta$ $\A$QCD, the coupling $\A(Q^2)$ differs from the underlying pQCD coupling $a(Q^2)$ nonnegligibly at high $|Q^2| > 1 \ {\rm GeV}^2$ [the index in Eq.~(\ref{diffAa}) is $N=1$ in (F)APT; $N=5$ in 2$\delta$ and 3$\delta$ $\A$QCD]. This implies that (F)APT has a certain ambiguity when ``normalizing'' the strength of $\A(Q^2)$. As mentioned, we fixed the strength of the coupling $\A(Q^2)$ in (F)APT to the value ${\overline \Lambda}_{N_f=3}=0.45$ GeV, since in such a case (F)APT reproduces approximately the QCD phenomenology at high energies (cf.~also \cite{Sh}). A question appears whether the results of our fits in (F)APT depend significantly on this value. We repeated the analysis in (F)APT with the value ${\overline \Lambda}_{N_f=3}=0.40$ GeV, and it turned out that the results of the fits did not change significantly. For example, when $Q^2_{\rm min}=0.268 \ {\rm GeV}^2$ and $\mu_6=0$, we obtained $k=2.86$ and the central value $f_2^{p-n}(1.)= -0.137$, and for the three quality parameters  $\chi^2_{\rm ext}/{\rm d.o.f.}$, $\chi^2_{0.268}/{\rm d.o.f.}$ and $\chi^2_{\rm all}/{\rm d.o.f}$ the values $120.$, $54.7$ and $878.$, respectively (to be compared with the corresponding values in Table \ref{tabmu60}, the third line from below). When $\mu_6 \not= 0$ in the fit, we obtained $k=2.62$, $f_2^{p-n}(1.)=-0.187$, $\mu_6=0.013 \ {\rm GeV}^4$ and for the mentioned three quality parameters the values $42.3$, $7.94$ and $4.43 \times 10^3$, respectively (to be compared with the corresponding values in Table \ref{tabmu6not0}, the third line from below).

\begin{figure}[htb] 
\begin{minipage}[t]{0.49\linewidth}
\centering\includegraphics[width=89mm]{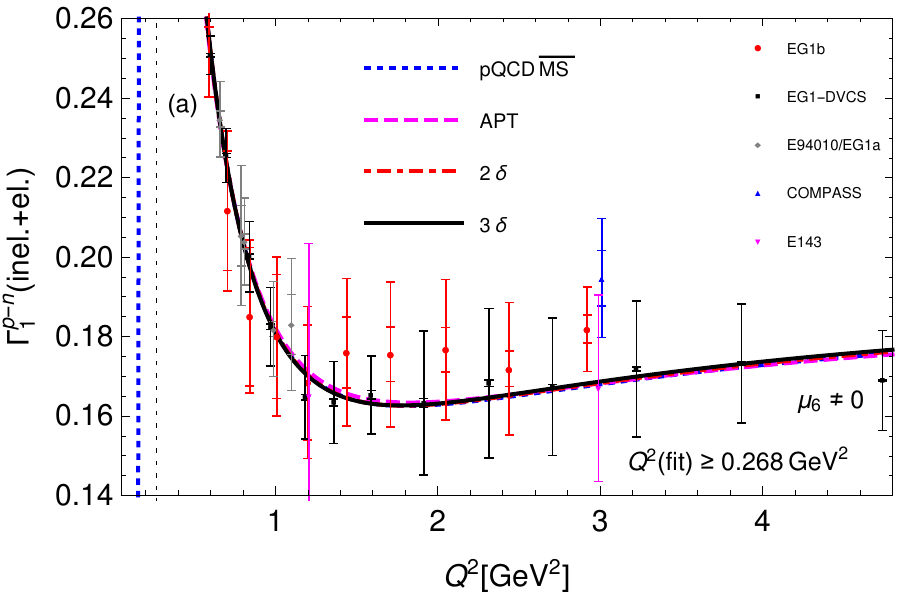}
\end{minipage}
\begin{minipage}[t]{0.49\linewidth}
\centering\includegraphics[width=89mm]{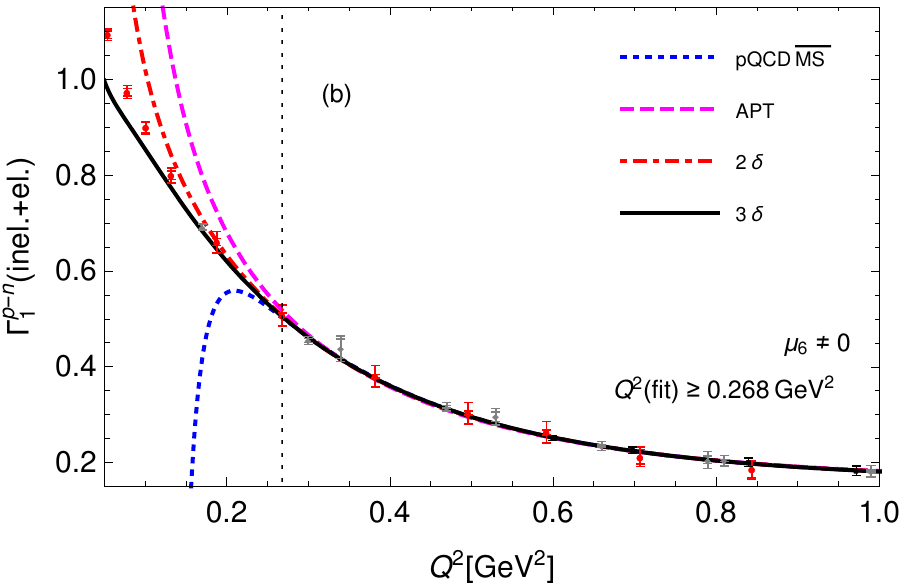}
\end{minipage}
\vspace{-0.4cm}
\caption{(color online): As Fig.~\ref{Fig0268}, but shifted upwards by the parametrized elastic contribution (\ref{G1el1}) [cf.~also Eqs.~(\ref{Ftau})-(\ref{Fj})].}
\label{Fig0268EL}
\end{figure}

When we add the parametrized elastic contribution of BSR, Eqs.~(\ref{G1el1})-(\ref{Fj}), to the experimental points and to the theoretical curves of Figs.~\ref{Fig0268}, we obtain the results presented in Figs.~\ref{Fig0268EL}(a),(b). In comparison with Figs.~\ref{Fig0268}, the values of BSR are shifted to significantly higher values at low $Q^2$. 
With this approach, the quality of fits ($\chi^2/{\rm d.o.f}$) does not change, as we consider the elastic contribution as known (and parametrized) and added here simultaneously to the (inelastic) theoretical fitting curves and to the data points. This means that the results of Table \ref{tabmu6not0} remain unchanged under this subsequent addition of $\Gamma_1^{p-n}(Q^2)_{\rm el.}$.

We can observe in the results of Tables \ref{tabmu6not0} and \ref{tabmu60} that the values of the higher-twist parameters, $f_2^{p-n}(1)$ and $\mu_6$, are in the analytic variants of QCD smaller than in pQCD, this reduction being especially strong in the 3$\delta$ QCD variant. It has been noted in the literature that in pQCD OPE there is a duality between the order of truncation of the leading-twist series and the higher-twist contribution \cite{PSTSK10,Kataev:1997nc,Narison:2009ag,Parente:1994bf}: higher-twist contribution often significantly decreases with the inclusion of higher orders in the leading-twist part. This effect and ambiguity become stronger in the ranges where the perturbation theory becomes questionable (for example, at the large and low values of the Bjorken variable $x$, as it was shown in Refs.~\cite{Kotikov:1992ht,Krivokhizhin:2005pt}, respectively). It has been observed that the higher-twist contribution is smaller, but also more stable (under the inclusion of more terms in the leading-twist), in QCD variants with infrared modifications of the coupling (various modifications lead to quite similar results \cite{Kotikov:2004uf}). The latter probably incorporate a part of the higher-twist contributions (which are rather cumbersome \cite{Illarionov:2004nw}) into (formally) the leading-twist contribution for small $x$ range at moderately small $Q^2$ values ($\lesssim 1 \ {\rm GeV}^2$) (see Ref.~\cite{Cvetic:2009kw} and more recent studies \cite{Kotikov:2012sm}  of the precise combined H1 and ZEUS data \cite{Aaron:2009aa} for the DIS structure function $F_2$).
  
Following the above observations in Tables \ref{tabmu6not0} and \ref{tabmu60}, we can conclude that the applications of the 2$\delta$ and especially 3$\delta$ $\A$QCD are very appropriate frameworks for the BSR studies because they appear to resum effectively a large part of the perturbative contribution into the leading-twist part (\ref{DBSAQCD}). In this context, we wish to recall that 3$\delta$ $\A$QCD is significantly different from the other two $\A$QCD variants [(F)APT and 2$\delta$ $\A$QCD] in the infrared region, because its coupling is not just finite there but goes to zero, $\A(Q^2) \sim Q^2 \to 0$, as motivated by large-volume lattice calculations, cf.~\cite{LattcoupNf0,LattcoupNf0b,LattcoupNf2,LattcoupNf4} and Appendix \ref{subsapp:2d3d}. Further, we wish to point out that the 2$\delta$ and 3$\delta$ $\A$QCD couplings $\A$ (and thus $\tA_n$ and $\A_n$) are at large $Q^2$ indistinguishable from their underlying pQCD couplings $a$ (${\widetilde a}_n$, $a^n$), cf.~Eq.~(\ref{diffAaN5}), in contrast to (F)APT which satisfies the relation Eq.~(\ref{diffAa}) with $N=1$. Therefore, theoretically, neither the higher-twist contribution of order $1/(Q^2)^N$ with $N \leq 4$ ($D \leq 8$), nor a part of it, is incorporated in the leading-twist contribution (\ref{DBSAQCD}) in the 2$\delta$ and 3$\delta$ $\A$QCD, in contrast to (F)APT. This indicates that the higher-twist terms extracted here with 2$\delta$ and 3$\delta$ $\A$QCD (in truncated OPE) represent an effective form for the true higher-twist contribution with dimension $D \leq 8$ [and a part of the other ($D \geq 10$) presumably small contribution]. In pQCD this is definitely not so, because of the mentioned duality there between the order of truncation of the leading-twist series and the extracted higher-twist contribution. As a consequence, the extracted effective higher-twist contribution in pQCD represents a sum of the true higher-twist contribution and a significant part of the perturbative (leading-twist) contribution; this effective higher-twist contribution appears to be in general larger than the true higher-twist contribution.

\subsection{Fits with ``massive'' OPE}
\label{subs:mass}

\begin{figure}[htb] 
\begin{minipage}[t]{0.49\linewidth}
\centering\includegraphics[width=89mm]{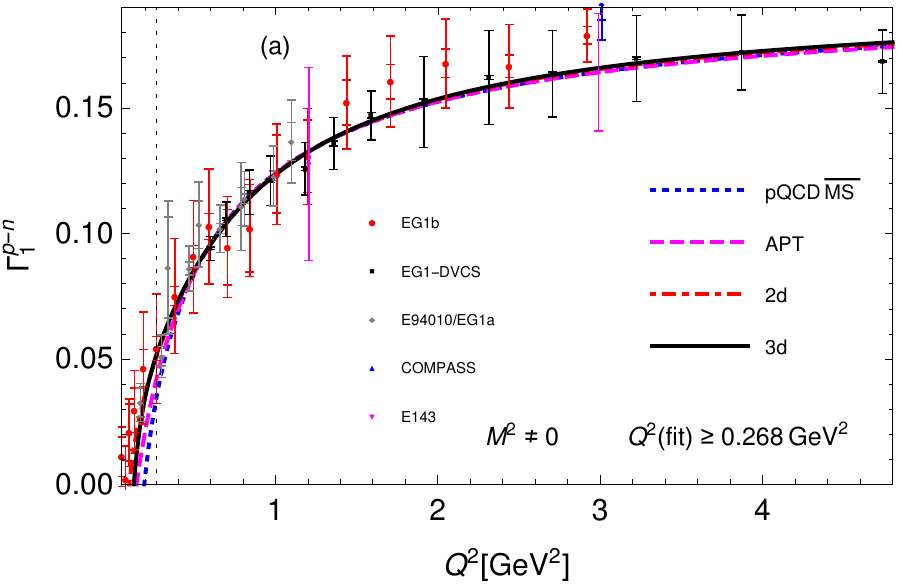}
\end{minipage}
\begin{minipage}[t]{0.49\linewidth}
\centering\includegraphics[width=89mm]{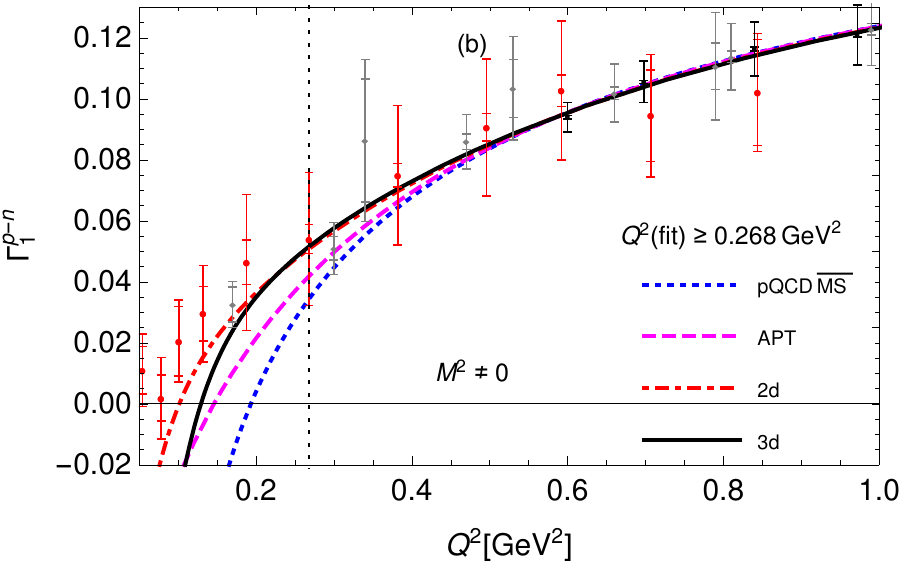}
\end{minipage}
\vspace{-0.4cm}
\caption{(color online): As Fig.~\ref{Fig0268}, but for the fit the OPE expression (\ref{BSROPEM}) with ``massive'' higher-twist term was used.}
\label{Fig0268M}
\end{figure}

For comparison, we performed a similar fit, but now with a ``massive'' higher-twist term instead of the truncated OPE expression (\ref{BSROPE})
\bea
\label{BSROPEM}
\Gamma_1^{p-n,{\rm mOPE[4]}}(Q^2; k, f_2^{p-n}(1); M^2) &=&
{\Big |}\frac{g_A}{g_V} {\Big |} \frac{1}{6}(1 - {\cal D}_{\rm BS}(Q^2)) +
\nonumber\\ &&
+ \frac{M_N^2}{(Q^2 + M^2)} \frac{1}{9} \left( a_2^{p-n} + 4 d_2^{p-n} + 4 f_2^{p-n}(Q^2) \right),
\eea
where the squared mass $M^2$ in the denominator of the higher-twist part\footnote{Similar higher-twist expressions were used in the analyses of BSR in \cite{KTG1,KTG2} where the leading-twist contribution was evaluated with the ``Massive'' Perturbation Theory (MPT) \cite{MPT}. MPT is an extension of APT \cite{ShS,MS96,ShS98,Sh} where, in contrast to APT, the coupling $\A(Q^2)$ is analytic in the point $Q^2=0$; nonetheless, the index $N$ of Eq.~(\ref{diffAa}) (Appendix \ref{app:An}) remains in MPT at the minimal value as in APT, i.e., $N=1$.}
is taken to be constant (not running), and is expected to be $0 < M^2 \lesssim 1 \ {\rm GeV}^2$. Now, instead of $f^{p-n}_2(1)$ and $\mu_6$, the fit parameters are $f_2^{p-n}(1)$ and $M^2$. The resulting curves, for $Q^2_{\rm min}=0.268 \ {\rm GeV}^2$, are given in Figs.~\ref{Fig0268M}(a),(b), at the higher $Q^2$ and the lower $Q^2 < 1 \ {\rm GeV}^2$ momenta, respectively.
The corresponding results are given in Table \ref{tabM}. These curves are analogous to those in the previous Figs.~\ref{Fig0268}(a),(b) in which the truncated OPE (\ref{BSROPE}) was used. Numerically, the behavior at low $Q^2$ in the ``massive'' case is significantly influenced by the $Q^2$-dependence of $f_2^{p-n}(Q^2)$. In the $\MSbar$ pQCD case, as in the $\MSbar$ pQCD cases of the analyses in all the Sections, we replaced in the higher-twist running parameter $f_2^{p-n}(Q^2)$ [cf.~Eq.~(\ref{HTQ2pt})] the scale $Q^2$ in an ad hoc way by the renormalization scale $k Q^2$ used in the leading-twist part ($k=16$ resulted here), in order to artificially avoid the problem of Landau singularities in the pQCD coupling $a(Q^2)$. Comparing Figs.~\ref{Fig0268M} and Table \ref{tabM} with the corresponding ``nonmassive'' case Figs.~\ref{Fig0268} and Table \ref{tabmu6not0}, we see that the results and extrapolations in the case of 3$\delta$ $\A$QCD are now comparably good in the ``massive'' and the $\mu_4 \& \mu_6$ approaches. Stated differently, the corresponding $\chi^2_{ \{... \}}/{\rm d.o.f.}$ values in Tables \ref{tabM} and \ref{tabmu6not0} (with $Q^2_{\rm fit}=0.268 \ {\rm GeV}^2$) are very similar. In the case of 2$\delta$ $\A$QCD and (F)APT, the extrapolations are better in the ``massive'' than in the $\mu_4 \& \mu_6$ approach, i.e., $\chi^2_{\rm all}/{\rm d.o.f.}$ is significantly reduced in the ``massive'' case. One reason for this lies perhaps in the fact that the massive higher-twist term is under control at very low $Q^2$, unlike the separate $\mu_4(Q^2)/Q^2$ and $\mu_6/(Q^2)^2$ terms. Further, the extracted values of $f_2^{p-n}(1)$ are in general similar in the $\mu_6=0$, $\mu_4 \& \mu_6$ and the ``massive'' approaches, although the uncertainties of the extracted parameters are quite high in the ``massive'' approach.\footnote{The systematic uncertainties of the extracted parameters in Table \ref{tabM} are large and should therefore be regarded as crude estimates only, cf.~comments in Appendix \ref{subsapp:mass}.} 
\begin{table}
  \caption{As in Table \ref{tabmu6not0}, but now the OPE form is ``massive'', Eq.~(\ref{BSROPEM}). The fits were made only for $Q^2_{\rm min}({\rm fit})=0.268 \ {\rm GeV}^2$. The extracted values of the squared mass $M^2$ are in ${\rm GeV}^2$.}
   \label{tabM}
\begin{ruledtabular}
\centering
\begin{tabular}{r l|llllll}
  QCD variant & $Q^2_{\rm min}({\rm fit})$ & $k$ & $f_2^{p-n}(1.)$ & $M^2$ & $\chi^2/{\rm d.o.f}$ & $\chi^2_{\rm ext}/{\rm d.o.f}$ & $\chi^2_{\rm all}/{\rm d.o.f.}$
\\
\hline
$\MSbar$ pQCD & 0.268 & 16.0 & $-0.286 \pm 0.003 \pm 1.035$ & $0.623 \pm 0.017 \pm 0.471$ & 7.94 & 35.4 & 130.
\\
(F)APT       & 0.268 & 16.0 & $-0.209 \pm 0.002 \pm 0.744$ & $0.439 \pm 0.012 \pm 0.463$ & 7.48 & 31.5 & 31.2 
\\
2$\delta$    & 0.268 & 0.689 & $-0.097 \pm 0.002 \pm 0.155$ & $0.336 \pm 0.023 \pm 0.371$ & 6.25 & 41.1 & 36.9
\\
3$\delta$ & 0.268 & 2.71 & $-0.065 \pm 0.001 \pm 0.462$ &  $0.036 \pm 0.017 \pm 0.377$ & 5.99 & 46.6 & 55.0
\end{tabular}
\end{ruledtabular}
\end{table}

The results of Table \ref{tabM} show that the QCD variants with infrared-finite analytic coupling, and especially 3$\delta$ $\A$QCD, give smaller values of higher-twist parameters $f_2^{p-n}(1)$ and $M^2$ than pQCD. On the one hand, at low $Q^2$, the smaller values of $M^2$ compensate partially the decreased value of $f_2^{p-n}$ in the higher-twist contribution. On the other hand, smaller values of $M^2$ and $f_2^{p-n}(1)$ mean that at higher values of $Q^2$ the higher-twist contribution is significantly reduced; this can be seen also by expanding the massive higher-twist term of Eq.~(\ref{BSROPEM}) in powers of $M^2/Q^2$. Such effect is in full agreement with one observed in Ref.\cite{PSTSK10}; the effect can be considered as a stabilization of the higher-twist contribution. We also recall that 3$\delta$ $\A$QCD is significantly different from (F)APT and 2$\delta$ $\A$QCD in the infrared region, since its coupling goes to zero there, $\A(Q^2) \sim Q^2 \to 0$.

It is possible to choose for the higher-twist term a massive form with a running mass, in the spirit of a dynamical effective gluon mass of the gluon propagator at low $Q^2$ \cite{Cornwall}. Such masses appear in the literature often in definitions of QCD couplings at low $Q^2$, and are responsible for the freezing (finiteness) of the coupling at $Q^2 \to 0$. The couplings in (F)APT and 2$\delta$ $\A$QCD have a freezing which could be described also via a running effective gluon mass. Our view is that the OPE higher-twist terms represent a new contribution not contained in the QCD coupling itself. The squared mass $M^2$ in such terms, Eq.~(\ref{BSROPEM}), is considered constant, in the spirit of the approach of Ref.~\cite{KTG1} (cf.~also \cite{KTG2}), where the basic component (delta function) in the spectral function of the higher-twist contribution gives such a mass term. Nonetheless, we repeated the aforementioned analysis for the case of a running squared mass $M^2(Q^2)$ representative of a dynamical effective gluon mass, chosen with a simple parametrization of Ref.~\cite{ABP}
\be
M^2(Q^2) = \frac{m_0^2}{1 + (Q^2/{\cal M})^{1+p}} \qquad ({\cal M}=0.5 \ {\rm GeV}, p=0.1), 
\label{Mv}
\ee
where we chose for the parameters ${\cal M}$ and $p$ values within the expected regions \cite{ABP}. The adjustable squared mass scale was taken (instead of $m_0^2$) to be $M^2(1 \ {\rm GeV}^2)$. The same analysis then gave the results presented in Table \ref{tabMv}. We can see that the results are qualitatively similar to those with the constant squared mass $M^2$, Table \ref{tabM}, except that the values of $M^2(1 \ {\rm GeV}^2)$ are now lower. Nonetheless, a reasonable definition of the average squared mass value in the present analysis, for the considered fit interval $0.268 \ {\rm GeV}^2 < Q^2 < 3 \ {\rm GeV}^2$ may be $\langle M^2(Q^2) \rangle = (1/2) \times ( M^2(0.268) + M^2(3) ) \approx 1.52 \times M^2(1 \ {\rm GeV}^2)$. Or, if we regard that the squared mass is most relevant only in the low-$Q^2$ part  $0.268 \ {\rm GeV}^2 < Q^2 < 1 \ {\rm GeV}^2$ of the fit interval, a reasonable definition of the average squared mass would be $\langle M^2(Q^2) \rangle = (1/2) \times ( M^2(0.268) + M^2(1) ) \approx 1.85 \times M^2(1 \ {\rm GeV}^2)$.
\begin{table}
  \caption{As in Table \ref{tabM}, but now the squared mass in the higher-twist term in Eq.~(\ref{BSROPEM}) is running according to Eq.~(\ref{Mv}).}
   \label{tabMv}
\begin{ruledtabular}
\centering
\begin{tabular}{r l|llllll}
  QCD variant & $Q^2_{\rm min}({\rm fit})$ & $k$ & $f_2^{p-n}(1.)$ & $M^2(1 \ {\rm GeV}^2)$ & $\chi^2/{\rm d.o.f}$ & $\chi^2_{\rm ext}/{\rm d.o.f}$ & $\chi^2_{\rm all}/{\rm d.o.f.}$
\\
\hline
$\MSbar$ pQCD & 0.268 & 11.9 & $-0.214 \pm 0.002 \pm 0.651$ & $0.240 \pm 0.010 \pm 0.272$ & 5.91 & 48.1 & 252.
\\
(F)APT       & 0.268 & 16.0 & $-0.171 \pm 0.002 \pm 0.186$ & $0.152 \pm 0.007 \pm 0.100$ & 6.65 & 43.2 & 38.8 
\\
2$\delta$    & 0.268 & 9.26 & $-0.160 \pm 0.002 \pm 0.496$ & $0.277 \pm 0.011 \pm 0.318$ & 5.88 & 49.0 & 48.0
\\
3$\delta$ & 0.268 & 2.65 & $-0.064 \pm 0.001 \pm 0.097$ &  $0.010 \pm 0.009 \pm 0.162$ & 5.99 & 46.5 & 48.3
\end{tabular}
\end{ruledtabular}
\end{table}

\subsection{Testing low-$Q^2$ regime ans\"atze}
\label{subs:lowQ2}

At low $Q^2$, the inelastic contribution to BSR behaves as $\sim Q^2$, according to Gerasimov-Drell-Hearn sum rule \cite{GDHsr} as pointed out and used in \cite{Ansel,PSTSK10,GDHlow}. Based on this, an expansion \cite{EG1b} motivated by chiral perturbation theory ($\chi$PT) can be constructed
\be
\Gamma_1^{p-n}(Q^2)_{\rm inel.} = \frac{\chi_n^2 - \chi_p^2}{8 M_N^2} Q^2 + A (Q^2)^2 + B (Q^2)^3 \qquad (Q^2 \lesssim 0.5 \ {\rm GeV}^2),
\label{chiPT}
\ee
where, according to Gerasimov-Drell-Hearn sum rule \cite{GDHsr}, $\chi_n$ and $\chi_p$ are anomalous magnetic moments of nucleons [which appear also in the elastic BSR contributions, cf.~Eqs.~(\ref{Fj})]; the parameters $A$ and $B$ are determined in the fit. When we fit with this expression the inelastic BSR data \cite{EG1a,EG1b,DVCS} for $Q^2 \leq Q^2_{\rm max}=0.2 \ {\rm GeV}^2$, we obtain $A=0.765 \ {\rm GeV}^{-2}$ and $B= 0.678 \ {\rm GeV}^{-4}$, with $\chi^2(Q^2 \leq Q^2_{\rm max})/{\rm d.o.f.} = 0.720$. On the other hand, if we take $Q^2_{\rm max}=0.5 \ {\rm GeV}^2$ in the fitting, we obtain  $A=0.744 \ {\rm GeV}^{-2}$ and $B= -1.033 \ {\rm GeV}^{-4}$, $\chi^2(Q^2 \leq Q^2_{\rm max})/{\rm d.o.f.} = 1.313$.

Another possible ansatz for the inelastic contribution to BSR at low $Q^2$ is the form of the light-front holographic (LFH) effective charge $\A^{\rm (LFH)}$ in the BSR ($g_1$) scheme \cite{LFH,LFHBSR} [$\A(0)_{g_1}=1$]
\be
\Gamma_1^{p-n}(Q^2)_{\rm inel.} = {\Big |} \frac{g_A}{g_V} {\Big |} \frac{1}{6} \left[ 1 - \A^{\rm (LFH)}(Q^2) \right]
=  {\Big |} \frac{g_A}{g_V} {\Big |} \frac{1}{6} \left[1 - \exp \left( - \frac{Q^2}{2 \kappa^2} \right) \right], \qquad (Q^2 \lesssim 1 \ {\rm GeV}^2).
\label{LFH}
\ee
When we fit with this expression the inelastic BSR data \cite{EG1a,EG1b,DVCS} for  $Q^2 < Q^2_{\rm max}=0.5 \ {\rm GeV}^2$, we obtain $\kappa=0.479$ GeV.\footnote{This is not far from the universal nonperturbative scale $\kappa= M_{\rho}/\sqrt{2} = 0.548$ GeV, cf.~\cite{LFHBSR}.}

\begin{figure}[htb] 
\begin{minipage}[t]{0.49\linewidth}
\centering\includegraphics[width=89mm]{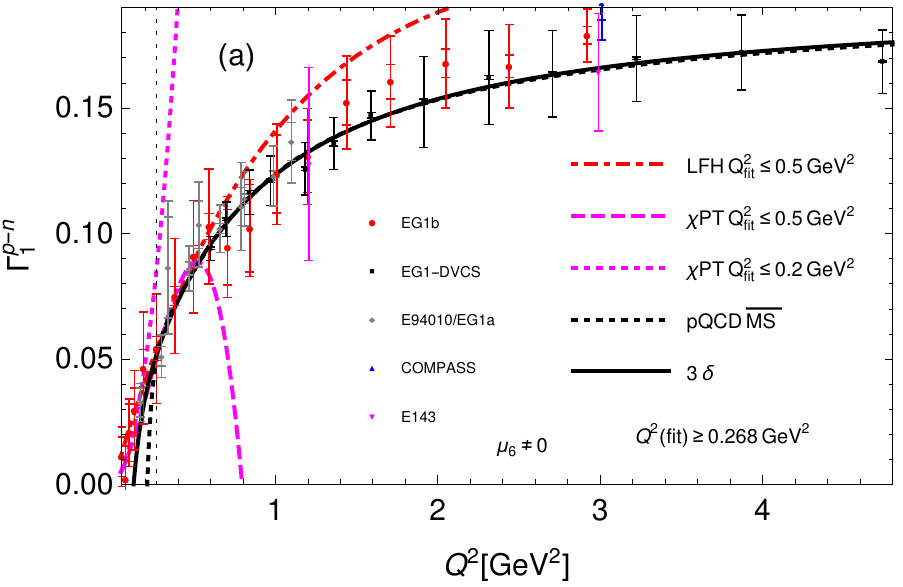}
\end{minipage}
\begin{minipage}[t]{0.49\linewidth}
\centering\includegraphics[width=89mm]{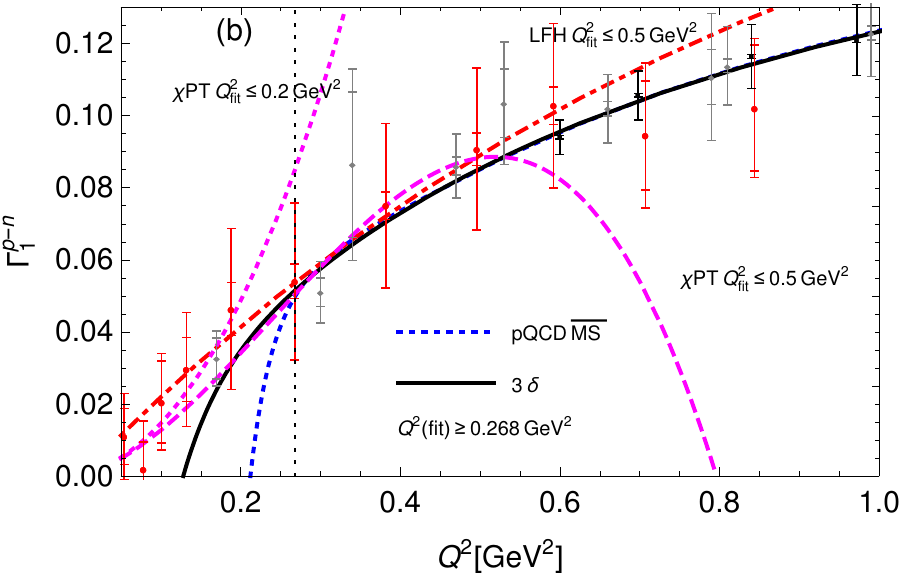}
\end{minipage}
\vspace{-0.4cm}
\caption{(color online): As Fig.~\ref{Fig0268}, but now with the low-$Q^2$ expressions (\ref{chiPT}) and (\ref{LFH}) included. Fig.~(b) is a zoomed-in version of Fig.~(a), for $Q^2 < 1 \ {\rm GeV}^2$. The lower bound of the fitting interval, $Q^2_{\rm min}=0.268 \ {\rm GeV}^2$, is included as the thin dotted vertical line.}
\label{Fig0268Q2Low}
\end{figure}

In Figs.~\ref{Fig0268Q2Low}(a),(b), we present these low-$Q^2$ expressions. The $\MSbar$ pQCD and 3$\delta$ $\A$QCD curves (obtained from fit with $Q^2 \geq Q^2_{\rm min} = 0.268 \ {\rm GeV}^2$) are also included for comparison.
In Fig.~\ref{Fig0268Q2Low}(b) we can see that the theoretical fitted curves of the presented QCD variants connect smoothly with the nonperturbative low-$Q^2$ curves ($\chi$PT and LFH, both fitted to BSR up to $Q^2=0.5 \ {\rm GeV}^2$) at the values of $Q^2$ around $0.3 \ {\rm GeV}^2$. Nonetheless, we recall that the apparent success of the ($\MSbar$) pQCD curve, down to about $0.3 \ {\rm GeV}^2$, was achieved due to the ad hoc change of scale in  $f^{p-n}_2(Q^2)$ to $k Q^2$ (with $k=3$), to avoid the Landau singularities in the $D=2$ term at $Q^2 \geq 0.268 \ {\rm GeV}^2$ (cf.~also the discussion about that point in Sec.~\ref{subs:mu6}). We further notice that the 3$\delta$ QCD curve agrees well with the mentioned $\chi$PT curve in a broader interval, $0.17 \ {\rm GeV}^2 < Q^2 < 0.3 \ {\rm GeV}^2$. Similar analyses with the goal of connecting the curves of pQCD (or of specific QCD variants) with nonperturbative curves at low $Q^2$ were performed in some of the references \cite{PSTSK10}, in \cite{LFHmatch}, and was discussed also in \cite{Burkert}.

In Refs.~\cite{Bernard,Lensky}, BSR at low $Q^2 < 0.3 \ {\rm GeV}^2$ was calculated with baryon chiral perturbation theory (B$\chi$PT) up to NLO. The authors of \cite{Bernard,Lensky} did not give the values of their parameters for BSR. However, careful visual comparison of their obtained $Q^2$-dependence [cf.~curves and bands in their Fig.~6(b) in Ref.~\cite{Lensky}] with our $\chi$PT curve (\ref{chiPT}) with $Q^2_{\rm fit} \leq 0.5 \ {\rm GeV}^2$ [the long-dashed curve in our Fig.~\ref{Fig0268Q2Low}(b)] shows very good agreement between them.
  
\subsection{Fitting to total (inelastic $+$ elastic) BSR data}
\label{subs:tot}

We perform also the fitting of the truncated (at $D=4$) OPE expression (\ref{BSROPE}) to the data for the total BSR $\Gamma_1^{p-n}(Q^2)_{\rm inel.+el.}$. As argued at the end of Sec.~\ref{subs:EC} [after Eq.~(\ref{G1elexp})], such a fit has problematic aspects. These data are obtained by adding to the experimental data $\Gamma_1^{p-n}(Q_j^2)_{\rm inel.}$ of Refs.~\cite{EG1a,EG1b,DVCS} the parametrized elastic contribution Eqs.~(\ref{G1el1})-(\ref{Fj}) of Ref.~\cite{Sabbiretal}. The uncertainties of the parametrized elastic form factors are considered to be less than $10 \%$ \cite{Sabbiretal}. Since the elastic contribution (\ref{G1el1}) is quadratic in the form factors (and is numerically dominated by the proton contribution), the relative uncertainties of the elastic contribution are less than $5 \%$. Since we do not have more information about these uncertainties, we will neglect them in this analysis. This means that in the total BSR, we will consider that the statistical and the systematic uncertainties $\sigma_{j, {\rm stat}}$ and  $\sigma_{j, {\rm sys}}$ are those of the inelastic contribution. Otherwise, the fitting is performed as in the previous Sec.~\ref{subs:mu6}.
\begin{figure}[htb] 
\begin{minipage}[t]{0.49\linewidth}
\centering\includegraphics[width=89mm]{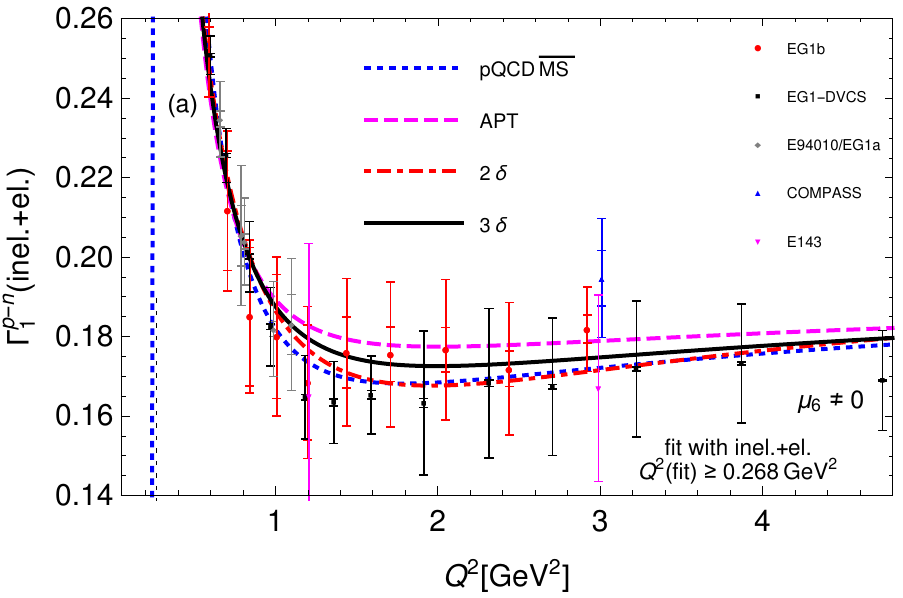}
\end{minipage}
\begin{minipage}[t]{0.49\linewidth}
\centering\includegraphics[width=89mm]{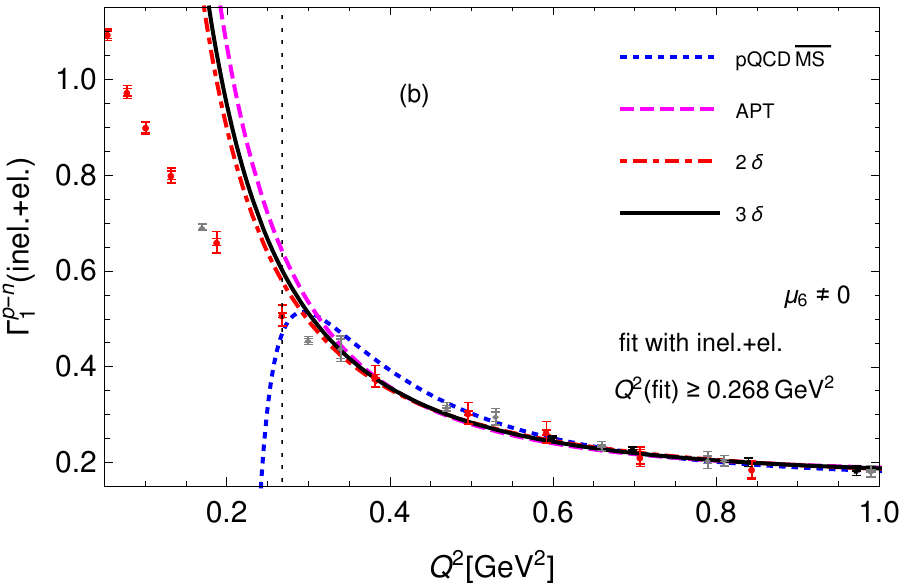}
\end{minipage}
\vspace{-0.4cm}
\caption{(color online): As Fig.~\ref{Fig0268}, but now the truncated (at $D=4$) OPE expression (\ref{BSROPE}) is fitted to the data for the total contribution $\Gamma_1^{p-n}(Q_j^2)_{\rm inel.+el.}$. The lower bound of the fitting interval, $Q^2_{\rm min}=0.268 \ {\rm GeV}^2$, is included as the thin dotted vertical line.}
\label{Fig0268fitwEL}
\end{figure}

The resulting curves are presented in Figs.~\ref{Fig0268fitwEL}(a),(b), for $Q^2 \geq Q^2_{\rm min}$ with $Q^2_{\rm min} = 0.268 \ {\rm GeV}^2$.
The obtained results are given in Table \ref{tabfitwEL}, for fits with various $Q^2_{\rm min} = 0.66$, $047$ and $0.268 \ {\rm GeV}^2$.
\begin{table}
\caption{As in Table \ref{tabmu6not0}, but now the fit is performed on the sum which includes the elastic contribution, i.e., on $\Gamma_1^{p-n}({\rm inel.+el.})$.}
\label{tabfitwEL}
\begin{ruledtabular}
\centering
\begin{tabular}{r l|lllllll}
  QCD variant & $Q^2_{\rm min}({\rm fit})$ & $k$ & $f_2^{p-n}(1.)$ & $\mu_6$ & $\chi^2/{\rm d.o.f}$ & $\chi^2_{\rm ext}/{\rm d.o.f}$ & $\chi^2_{0.268}/{\rm d.o.f.}$ & $\chi^2_{\rm all}/{\rm d.o.f.}$
\\
\hline
$\MSbar$ pQCD & 0.66 & 16. & $-0.253 \pm 0.002 \pm 0.125$ & $0.101 \pm 0.001 \pm 0.064$ & 7.11 & 55.6 & $3.65 \times 10^5$ & $1.26 \times 10^{6}$
\\
(F)APT       & 0.66 & 16. & $-0.227 \pm 0.002 \pm 0.105$ & $0.086 \pm 0.001 \pm 0.028$ & 7.53 & 46.3 & 697. & $3.54 \times 10^{5}$
\\
2$\delta$    & 0.66 & 0.487 & $-0.121 \pm 0.002 \pm 0.057$ & $0.073 \pm 0.001 \pm 0.011$ & 7.18 & 52.0 & 555. & $2.60 \times 10^{5}$
\\
3$\delta$ & 0.66 & 2.01 & $-0.095 \pm 0.002 \pm 0.145$ &  $0.067 \pm 0.001 \pm 0.033$ &7.04 & 55.8 & 509. & $2.27 \times 10^{5}$
\\
\hline
$\MSbar$ pQCD & 0.47 & 1.37 & $-0.194 \pm 0.002 \pm 0.111$ & $0.114 \pm 0.001 \pm 0.056$ & 8.82 & 39.0 & $\infty$ & $\infty$
\\
(F)APT       & 0.47 & 1.67 & $-0.191 \pm 0.002 \pm 0.086$ & $0.072 \pm 0.001 \pm 0.021$ & 19.4 & 63.4 & 364. & $2.46 \times 10^{5}$
\\
2$\delta$    & 0.47 & 0.194 & $-0.089 \pm 0.002 \pm 0.052$ & $0.059 \pm 0.001 \pm 0.093$ & 14.8 & 57.8 & 250. & $1.67 \times 10^{5}$
\\
3$\delta$ & 0.47 & 0.775 & $-0.076 \pm 0.002 \pm 0.135$ &  $0.053 \pm 0.001 \pm 0.029$ & 12.6 & 44.9 & 193. & $1.38 \times 10^{5}$
\\
\hline
$\MSbar$ pQCD & 0.268 & 1.82 & $-0.153 \pm 0.001 \pm 0.081$ & $0.086 \pm 0.000 \pm 0.010$ & 28.0 & 82.3 & 28.0 & $\infty$
\\
(F)APT       & 0.268 & 1.98 & $-0.096 \pm 0.001 \pm 0.054$ & $0.044 \pm 0.000 \pm 0.011$ & 131. & 254. & 131. & $8.84 \times 10^{4}$
\\
2$\delta$    & 0.268 & 0.063 & $0.018 \pm 0.001 \pm 0.042$ & $0.028 \pm 0.000 \pm 0.006$ & 37.9 & 113. & 37.9 & $3.91 \times 10^{4}$
\\
3$\delta$ & 0.268 & 0.846 & $0.000 \pm 0.001 \pm 0.091$ &  $0.030 \pm 0.000 \pm 0.016$ & 68.7 & 145. & 68.7 & $4.61 \times 10^{4}$
\end{tabular}
\end{ruledtabular}
\end{table}
Comparing the obtained results in Figs.~\ref{Fig0268fitwEL} and Table \ref{tabfitwEL} with the corresponding results in Figs.~\ref{Fig0268EL} and Table \ref{tabmu6not0} (where the elastic part was not included in the fit procedure), we see that the inclusion of the elastic contribution in the fit procedure significantly deteriorates (increases) the values of the various fit quality parameters $\chi^2$, especially when the fit is performed in the larger interval $(Q^2_{\rm min}=) 0.268 \ {\rm GeV}^2 \leq Q^2 < 3 {\rm GeV}^2$. In particular, the differences in quality are clearly visible when comparing Fig.~\ref{Fig0268fitwEL}(b) with Fig.~\ref{Fig0268EL}(b) at low $Q^2$.
  Only when $Q^2_{\rm min}$ is relatively high, $Q^2_{\rm min}=0.66 \ {\rm GeV}^2$, are some of the $\chi^2$ parameters comparable in the two cases (but not the extrapolation quality parameters $\chi^2_{0.268}$ and $\chi^2_{\rm all}$). Further, the inclusion of the elastic contribution in the fit in general does not change significantly the extracted (mostly negative) values of $f_2^{p-n}(1 {\rm GeV}^2)$, but increases significantly the (positive) values of $\mu_6$. 

For the fit analysis with the ``massive'' truncated OPE, Eq.~(\ref{BSROPEM}), the results are presented in Table \ref{tabMfitwEL} for $Q^2_{\rm min}=0.268 \ {\rm GeV}^2$.
\begin{table}
  \caption{As in Table \ref{tabM}, i.e., the OPE form is ``massive'', Eq.~(\ref{BSROPEM}), but now the elastic part of BSR is included in the fit.}
   \label{tabMfitwEL}
\begin{ruledtabular}
\centering
\begin{tabular}{r l|llllll}
  QCD variant & $Q^2_{\rm min}({\rm fit})$ & $k$ & $f_2^{p-n}(1.)$ & $M^2$ & $\chi^2/{\rm d.o.f}$ & $\chi^2_{\rm ext}/{\rm d.o.f}$ & $\chi^2_{\rm all}/{\rm d.o.f.}$
\\
\hline
$\MSbar$ pQCD & 0.268 & 1.47 & $0.117 \pm 0.000 \pm 0.042$ & $-0.247 \pm 0.000 \pm 0.017$ & 384. &  795. & $\infty$
\\
(F)APT       & 0.268 & 3.39 & $0.041 \pm 0.001 \pm 0.032$ & $-0.207 \pm 0.002 \pm 0.063$ & 395. & 760. & $4.63 \times 10^{3}$ 
\\
2$\delta$    & 0.268 & 0.063 & $0.086 \pm 0.001 \pm 0.044$ & $-0.156 \pm 0.002 \pm 0.032$ & 96.6 & 245. & $1.39 \times 10^{4}$
\\
3$\delta$ & 0.268 & 1.13 & $0.083 \pm 0.001 \pm 0.082$ &  $-0.151 \pm 0.003 \pm 0.071$ & 143. & 320. & $1.05 \times 10^{4}$
\end{tabular}
\end{ruledtabular}
\end{table}
In general, the results are worse than in the corresponding ``massive'' OPE case fitted to the inelastic contribution (cf.~Table \ref{tabM} in Sec.~\ref{subs:mu6}). Namely, the extracted values of the parameter $M^2$ become negative ($M^2 \sim -0.1 \ {\rm GeV}^2$) and the fit quality values $\chi^2/{\rm d.o.f.}$ are in general significantly higher than those in Table \ref{tabM}. The uncertainties of the extracted values of $f_2^{p-n}(1)$ and $M^2$ become reduced (in comparison to the case when the elastic part is not included in the fit); this is so because the elastic parts increase significantly BSR $\Gamma_1^{p-n}(Q^2)$ at low $Q^2$ where they are represented mostly by the ``massive'' higher-twist term, and hence the relative uncertainties of BSR at low $Q^2$ become smaller.

In addition to the massive case with constant squared mass $M^2$, we include also the case of $Q^2$-dependent squared mass Eq.~(\ref{Mv}) in the analysis with the elastic contribution included in the fit. The results for this case are presented in Table \ref{tabMvfitwEL}. These results are qualitatively similar to those with constant squared mass, Table \ref{tabMfitwEL}; and the comparison of Table \ref{tabMvfitwEL} with its counterpart Table \ref{tabMv} without the elastic contribution is similar to the above comparison when the squared mass is constant. We recall that an average mass in the $Q^2$-independent case can be regarded to be $\langle M^2(Q^2) \rangle = (1/2) \times ( M^2(0.268) + M^2(1) ) \approx 1.85 \times M^2(1 \ {\rm GeV}^2)$. 
\begin{table}
  \caption{As in Table \ref{tabMfitwEL}, but with $Q^2$-dependent mass Eq.~(\ref{Mv}) in the higher-twist part of Eq.~(\ref{BSROPEM}).}
    \label{tabMvfitwEL}
\begin{ruledtabular}
\centering
\begin{tabular}{r l|llllll}
  QCD variant & $Q^2_{\rm min}({\rm fit})$ & $k$ & $f_2^{p-n}(1.)$ & $M^2(1 \ {\rm GeV}^2)$ & $\chi^2/{\rm d.o.f}$ & $\chi^2_{\rm ext}/{\rm d.o.f}$ & $\chi^2_{\rm all}/{\rm d.o.f.}$
\\
\hline
$\MSbar$ pQCD & 0.268 & 4.88 & $0.113 \pm 0.001 \pm 0.053$ & $-0.065 \pm 0.001 \pm 0.070$ & 461. &  932. & $\infty$
\\
(F)APT       & 0.268 & 3.22 & $0.052 \pm 0.001 \pm 0.030$ & $-0.072 \pm 0.001 \pm 0.034$ & 436. & 825. & $3.25 \times 10^{3}$ 
\\
2$\delta$    & 0.268 & 0.063 & $0.102 \pm 0.001 \pm 0.034$ & $-0.051 \pm 0.001 \pm 0.020$ & 119. & 288. & $6.58 \times 10^{5}$
\\
3$\delta$ & 0.268 & 1.14 & $0.098 \pm 0.001 \pm 0.044$ &  $-0.048 \pm 0.001 \pm 0.023$ & 167. & 366. & $5.49 \times 10^{4}$
\end{tabular}
\end{ruledtabular}
\end{table}

\section{Summary}
\label{sec:concl}

Experimental results for the polarized Bjorken sum rule (BSR) $\Gamma_1^{p-n}(Q^2)$ were fitted, for various ranges of $Q^2$, with OPE theoretical expressions using QCD couplings obtained in four different approaches: perturbative QCD (pQCD) in $\MSbar$ scheme; (Fractional) Analytic Perturbation Theory [(F)APT]; Two-delta $\A$QCD (2$\delta$); and Three-delta lattice-motivated $\A$QCD (3$\delta$). The QCD running coupling $\A(Q^2)$ in the latter three QCD variants does not have Landau singularities, in contrast to the pQCD coupling $a(Q^2)$ [$\equiv \alpha_s(Q^2)/\pi$].

In the fit of the inelastic experimental BSR results, up to two higher-twist terms [$\sim 1/Q^2, 1/(Q^2)^2$] were added to the theoretical leading-twist contribution. The elastic contributions, \textcolor{black}{which are $\sim 1/(Q^2)^n$ with typically $n \geq 4$} at $Q^2 > 1 \ {\rm GeV}^2$, were then added by using the parametrization obtained from the literature. The fits were performed for the ranges $Q^2_{\rm min} \leq Q^2 \leq 3 \ {\rm GeV}^2$, where $Q^2_{\rm min} = 0.66$, $0.47$ and $0.268 \ {\rm GeV}^2$. In general, the best curves were obtained when 2$\delta$ or 3$\delta$-couplings were used. When only $D=2$ ($\sim 1/Q^2$) higher-twist term was included in the fit, the quality of the fitted curves, in the range of the fit and in the extrapolated ranges of $Q^2$, in general did not depend significantly on $Q^2_{\rm min}$ of the fit. On the other hand, when both $D=2$ and $D=4$ terms were included, the quality in the extrapolated ranges of $Q^2$ was in general better for the lowest $Q^2_{\rm min}$ value ($0.268 \ {\rm GeV}^2$), i.e., when the $Q^2$-range of the fit was the largest. Comparably good results were obtained when ``massive'' higher-twist term was used in the OPE and the QCD coupling was either from (F)APT or 2$\delta$ or 3$\delta$ $\A$QCD.

When the range of fit had $Q^2_{\rm min}=0.268 \ {\rm GeV}^2$, the pQCD $\MSbar$ coupling approach worked and gave acceptable results only if the renormalization scale of the coupling was maintained everywhere at sufficiently high values, and the coefficient $f_2^{p-n}(Q^2)$ [$\sim a(Q^2)^{\gamma_0/8 \beta_0}$] at the $D=2$ term had an (ad hoc) increased scale $Q^2 \mapsto k Q^2$,  in order to avoid the problem of the Landau singularities.

When the fit procedure was performed by fitting the theoretical OPE, truncated at $D=4$ ($\sim  1/(Q^2)^2$) terms, to the sum of (experimental) inelastic and (parametrized) elastic BSR, the quality of the results turned out to be significantly worse in all the cases of the theoretical curves, something expected by the arguments presented at the end of Sec.~\ref{subs:EC}.
Namely, the elastic contribution is dominated by terms which behave at high values of $Q^2$ as $\sim 1/(Q^2)^{(D/2)}$ where usually $D \geq 8$, and these terms are not contained in the theoretical expressions for BSR which are usually OPE series truncated at $1/(Q^2)^2$.

The results of this work can be interpreted as an additional indication of the following important property: the evaluation of the (truncated) leading-twist contribution of spacelike low-$Q^2$ QCD observables such as inelastic BSR, in QCD variants 2$\delta$ and in particular 3$\delta$ $\A$QCD [both have infrared finite and holomorphic coupling $\A(Q^2)$], appear to resum effectively a large part of the perturbative contribution of the observables, and leads to reduced extracted values of the higher-twist terms ($D=2, 4$) in the truncated OPE. This property was noted earlier, for different observables, in Refs.~\cite{anOPE,3l3danQCD,4l3danQCD}. In this context, it appears to be important that in 2$\delta$ and 3$\delta$ $\A$QCD the coupling practically merges with the underlying pQCD coupling $a(Q^2)$ at higher values of $Q^2 \gg \Lambda^2_{\rm QCD}$. This property is not shared by the (F)APT holomorphic coupling where the leading-twist series contains parts of the higher-twist contribution of as low dimensionality as $D=2$. The extracted parameters in the higher-twist contribution, including those in the ``massive'' OPE, are especially reduced in 3$\delta$ $\A$QCD. This suggests the possibility that the true higher-twist contribution is small, including the (sum of) terms of high dimension; and that the (truncated) OPE with 3$\delta$ $\A$QCD leading-twist gives, through fitting, an extracted value which is a good approximation to this true value of the higher-twist contribution. Numerically, the significantly reduced extracted value in (truncated) OPE with 3$\delta$ $\A$QCD is probably partly related with the fact that 3$\delta$ $\A$QCD differs from both 2$\delta$ and (F)APT $\A$QCD variants in that its coupling goes to zero in the deep infrared regime, $\A^{(3 \delta)}(Q^2) \sim Q^2 \to 0$. The latter property, we recall, is suggested by the large-volume lattice calculations of the dressing functions of the Landau-gauge gluon and ghost propagators at low $Q^2$ values.

\begin{acknowledgments}
\noindent
This work was supported by FONDECYT Postdoctoral Grant No.~3170116 (C.A.), by FONDECYT Regular Grant No.~1180344 (G.C. and C.A.), and by the RFBR Foundation through Grant No. 16-02-00790-a (A.V.K. and B.G.S.). We thank A.~Deur, J.~Bl\"umlein and A.~L.~Kataev for helpful suggestions.
\end{acknowledgments}

\appendix

\section{Renormalization scale and scheme dependence of the expansion coefficients}
\label{app:d1d2d3}

The dependence of the coupling $a(\mu^2) \equiv \alpha_s(\mu^2)/\pi$ (where $\mu^2>0$ means the spacelike region) is governed by the perturbative renormalization group equation (RGE)
\be
\frac{\partial}{\partial \ln \mu^2} a(\mu^2) = - \beta_0 a(\mu^2)^2 - \beta_1 a(\mu^2)^3 - \beta_2 a(\mu^2)^4 - \ldots,
\label{pRGE}
\ee
where in the mass-independent schemes the coefficients $\beta_0 = (11 - 2 N_f/3)/4$ and $\beta_1 = (102 - 38 N_f/3)/16$ are universal (scheme-independent), while the coefficients $\beta_j$ (or equivalently, $c_j \equiv \beta_j/\beta_0$) for $j \geq 2$ are the (arbitrary) parameters which characterize the renormalization scheme.\footnote{There is another scheme parameter, the scale $\Lambda^2$, such that $a(\mu^2) = f(\mu^2/\Lambda^2)$. However, the change of $\Lambda^2$ can be regarded as the change in the definition of the renormalization scale $\mu^2$.} The running of the coupling $a(\mu^2; c_2, c_3,\ldots)$ with these scheme parameters is governed by the following relations (cf.~App.~A of Ref.~\cite{Stevenson}, and App.~A of Ref.~\cite{GCRK63}):
\be
\frac{\partial a}{\partial c_2} = a^3 + {\cal O}(a^5), \qquad
\frac{\partial a}{\partial c_3} = \frac{1}{2} a^4 + {\cal O}(a^5), \qquad
\frac{\partial a}{\partial c_4} = {\cal O}(a^5),
\ldots
\label{aRSch}
\ee
When we use the relations (\ref{pRGE})-(\ref{aRSch}) in the perturbation expansion (\ref{DBSpt}) and account for the fact that ${\cal D}_{\rm BSR}(Q^2)$ is a (spacelike) observable and thus independent of the scale $\mu^2$ (i.e., independent of $k \equiv \mu^2/Q^2$) and of the scheme parameters $c_j$ ($j \geq 2$), we obtain the following expressions for the perturbation coefficients $d_j$ in terms of the general renormalization scale and scheme parameters ($k, c_2, c_3$):
\bes
\label{d1d2d3}
\bea
d_1(k) &=& {\bar d}_1 + \beta_0 \ln k,
\label{d1}
\\
d_2(k;c_2) &=& {\bar d}_2 + {\bar d}_1 2 \beta_0 \ln k + \beta_0^2 \ln^2 k + \beta_0 c_1 \ln k - (c_2 - {\bar c}_2);
\label{d2}
\\
d_3(k;c_2,c_3) &=& \bigg\{ \left[ {\bar d}_3 + {\bar d}_2 (3 \beta_0 \ln k) + {\bar d}_1 \left( 3 \beta_0^2 \ln^2 k + 2 \beta_0 c_1 \ln k \right) + \left (\beta_0^3 \ln^3 k + \frac{5}{2} \beta_0^2 c_1 \ln^2 k + \beta_0 {\bar c}_2 \ln k \right) \right]
\nonumber\\
&&
- 2 (c_2 - {\bar c}_2) ( {\bar d}_1 + \beta_0 \ln k) 
- \frac{1}{2}(c_3 - {\bar c}_3) \bigg\}.
\label{d3}
\eea
\ees
Here we used the bar symbol to denote the choice of the scheme $\MSbar$ with the renormalization scale $\mu^2=Q^2$ ($k=1$).
We recall that $k \equiv \mu^2/Q^2$ is the renormalization scale parameter ($0 < k \sim 1$), and that ${\bar c}_j \equiv {\bar \beta}_j/\beta_0$ ($j \geq 2$) are the $\MSbar$ scheme parameters.

\section{Power analogs $\A_n$ in $\A$QCD}
\label{app:An}

A QCD running coupling $\A(Q^2)$ which is holomorphic (analytic) in the non-timelike $Q^2$ complex plane sector $(-q^2 \equiv) Q^2 \in \mathbb{C} \backslash (-\infty, -M_{\rm thr}^2]$ (where $0 \leq M_{\rm thr}^2 \lesssim 1 \ {\rm GeV}^2$), has in general nonperturbative (NP) contributions $\sim 1/(Q^2)^n$ appreciable at small $|Q^2|$. It differs from the underlying pQCD coupling $a(Q^2)$ by
\be
\A(Q^2) - a(Q^2) \sim \left( \frac{\Lambda^2}{Q^2} \right)^N,
\label{diffAa}
\ee
for $|Q^2| > \Lambda^2$ ($\gtrsim 0.1 \ {\rm GeV}^2$). Here, $N=1$ in the case of (F)APT, and $N=5$ in 2$\delta$ and 3$\delta$ $\A$QCD. The analytization procedure can be presented schematically as $a(Q^2) \mapsto \A(Q^2)$. This procedure involves $a$ and $\A$ linearly (not as powers). Namely, when $Q^2$ is varied, $Q^2 \mapsto Q^2 + \Delta Q^2$, we have $a(Q^2 + \Delta Q^2) \mapsto \A(Q^2 + \Delta Q^2)$. Therefore,
\be
{\ta}_{n}(Q^2) \mapsto {\tA}_{n}(Q^2),
\label{tatA}
\ee
where we denoted the (logarithmic) derivatives
\bes
\label{tantAn}
\bea
{\ta}_{n}(Q^2) & \equiv & \frac{(-1)^{n-1}}{\beta_0^{n-1} (n-1)!} \left( Q^2 \frac{d}{d Q^2} \right)^{n-1} a(Q^2),
\label{tan}
\\
{\tA}_{n}(Q^2) &\equiv&  \frac{(-1)^{n-1}}{\beta_0^{n-1} (n-1)!} \left( Q^2 \frac{d}{d Q^2} \right)^{n-1} \A(Q^2), \qquad (n=1,2,\ldots).
\label{tAn}
\eea
\ees
In this notation, ${\ta}_1 = a$ and ${\tA}_1 \equiv \A$. We note that by pQCD RGE (\ref{pRGE}) we have
\be
{\ta}_{n}(Q^2)  =  a(Q^2)^{n} + {\cal O}(a^{n+1}).
\label{tan1}
\ee
More specifically, we have
\bes
\label{tans}
\bea
 {\ta}_2 &=& a^2 + c_1 a^3 + c_2 a^4 + \ldots,
 \label{ta2}
 \\
 {\ta}_3 & = & a^3 + \frac{5}{2} c_1 a^4 + \ldots, \qquad
 {\ta}_4 = a^4 + \ldots, \quad {\rm etc.},
 \label{ta3ta4}
 \eea
 \ees
 where, as mentioned in Appendix \ref{app:d1d2d3}, $c_j \equiv \beta_j/\beta_0$.
 When we invert these relations, we obtain
 \bes
 \label{ans}
 \bea
 a^2 & = & {\ta}_2 - c_1 {\ta}_3 + \left( \frac{5}{2} c_1^2 - c_2 \right) {\ta}_4 + \ldots,
\label{a2}
\\
a^3 & = & {\ta}_3 - \frac{5}{2} c_1 {\ta}_4 + \ldots, \qquad a^4 = {\ta}_4 + \ldots, \quad {\rm etc.}
\label{a3a4}
\eea
\ees
The linearity of analytization, Eq.~(\ref{tatA}), then gives us the analogs $\A_{n}$ of the powers $a^{n}$
\bes
 \label{Ans}
 \bea
 \A_2 & = & {\tA}_2 - c_1 {\tA}_3 + \left( \frac{5}{2} c_1^2 - c_2 \right) {\tA}_4 + \ldots,
\label{A2}
\\
\A_3 & = & {\tA}_3 - \frac{5}{2} c_1 {\tA}_4 + \ldots, \qquad \A_4 = {\tA}_4 + \ldots, \quad {\rm etc.}
\label{A3A4}
\eea
\ees
We note that in general $\A_n(Q^2) \not= \A(Q^2)^{n}$. The described construction (for $n=1,2,3,\ldots$) was performed in \cite{CV1,CV2}.

The above approach was extended in Ref.~\cite{GCAK} to the case of general real index $n = \nu$
\be
\tA_{\nu}(Q^2) = \frac{1}{\pi} \frac{(-1)}{\beta_0^{\nu-1} \Gamma(\nu)}
\int_{0}^{\infty} \ \frac{d \sigma}{\sigma} \rho_{\A}(\sigma)  
{\rm Li}_{-\nu+1}\left( - \frac{\sigma}{Q^2} \right) \quad (0 < \nu) \ ,
\label{tAnu}
\ee
where ${\rm Li}_{-\nu+1}(z)$ is the polylogarithm function of order $-\nu+1$, and $\rho_{\A}(\sigma) = {\rm Im} \A(Q^2=-\sigma - i \epsilon)$ is the cut discontinuity (spectral) function of $\A$. The coupling ${\tA}_{\nu}$ can also be presented in an alternative form applicable in an extended region $-1 < \nu$ (cf.~\cite{GCAK} for details). The expression $\A_{\nu}$, the analog of the power $a^{\nu}$, was then obtained in the form
\be
\A_{\nu}(Q^2) \equiv {\tA}_{\nu}(Q^2) + \sum_{m=1,2,\ldots}
\tk_m(\nu) {\tA}_{\nu + m}(Q^2) \quad (-1 < \nu) \ ,
\label{AnutAnu}
\ee
with the coefficients $\tk_m(\nu)$ given in Appendix A of Ref.~\cite{GCAK}. Eqs.~(\ref{Ans}) are a special case of Eq.~(\ref{AnutAnu}).

The perturbation expansion of the type Eq.~(\ref{DBSpt}) in Sec.~\ref{subs:PT}, for any spacelike obervable ${\cal D}(Q^2)$ in pQCD, can be reexpressed in terms of derivatives ${\ta}_{n}$ of Eq.~(\ref{tan})
\be
{\cal D}(Q^2)_{{\rm mpt}} = a + {\widetilde d}_1 {\ta}_2 + {\widetilde d}_2 {\ta}_3 + {\widetilde d}_3 {\ta}_4 + {\cal O}({\ta}_5),
\label{Dmpt}
\ee
where we denoted $a \equiv a(k Q^2; c_2,\ldots)$, ${\ta}_{n} \equiv {\ta}_{n}(k Q^2; c_2,\ldots)$, and the coefficients ${\widetilde d}_n \equiv {\widetilde d}_n (k; c_2,\ldots,c_n)$ of this ``modified'' perturbation expansion (mpt) can be obtained by using the RGE-relations (\ref{ans})
\bes
\label{td}
\bea
{\tilde d}_1 & = & d_1, \qquad {\tilde d}_2 = d_2 - c_1 d_1,
\label{td1td2}
\\
{\tilde d}_3 & = & d_3 - \frac{5}{2} c_1 d_2 + \left( \frac{5}{2} c_1^2 - c_2 \right) d_1, \qquad {\rm etc.},
\label{td3}
\eea
\ees
and the coefficients $d_j \equiv d_j(k; c_2,\ldots,c_j)$ are those of Eqs.~(\ref{d1d2d3}).
The expressions in $\A$QCD, corresponding to the perturbation expansions (\ref{Dmpt}) and (\ref{DBSpt}), are then
\bes
\label{DAQCD}
\bea
{\cal D}(Q^2)_{{\A}{\rm QCD}} &=&  \A + {\widetilde d}_1 {\tA}_2 + {\widetilde d}_2 {\tA}_3 + {\widetilde d}_3 {\tA}_4 + {\cal O}({\tA}_5),
\label{mAQCD}
\\
& = & \A + d_1 \A_2 + d_2 \A_3 + d_3 \A_4 + {\cal O}(\A_5).
\label{AQCD}
\eea
\ees
The expansion (\ref{AQCD}) is written again, in a more detailed form, in Eq.~(\ref{DBSAQCD}) in Sec.~\ref{subs:PT}. Both expressions (\ref{DAQCD}) are equivalent, but in practice it is more economical to do numerical evaluations using the expression (\ref{mAQCD}).

We have $\A_n(Q^2) = \A(Q^2)^n$ only when $\A(Q^2)$ is a perturbative coupling, i.e., when it has no NP terms ($\sim 1/(Q^2)^m$). It is important not to use the power expansion in $\A$ for the evaluation of spacelike observables. Namely, if we used power expansion in $\A$, the truncated series for ${\cal D}(Q^2)$ would have increasingly large (out of control) NP contributions when more power terms were included, and renormalization scale invariance would be increasingly violated, as emphasized in \cite{Techn}. It turns out that in practice the sequence ${\tA}_n(Q^2)$ ($n=0,1,2,\ldots$) is, in a general holomorphic $\A$QCD, a sequence with decreasing absolute values, at any finite $Q^2$: $|{\tA}_{n}(Q^2)| > |{\tA}_{n+1}(Q^2)| > \ldots$. In pQCD (${\ta}_n(Q^2)$) this is in general not valid at low values $|Q^2| \lesssim 1 \ {\rm GeV}^2$.

\section{QCD variants with holomorphic coupling $\A(Q^2)$}
\label{app:holA}

\subsection{(Fractional) Analytic Perturbation Theory [(F)APT]}
\label{subsapp:FAPT}

The pQCD running coupling $a(Q^2)$, in a given renormalization scheme (usually $\MSbar$), has in the complex $Q^2$-plane cut along the real axis, $(-\infty,\Lambda_{\rm Lan.}^2)$, where $0 < \Lambda_{\rm Lan.}^2 \sim 0.1 \ {\rm GeV}^2$ is the branching point of the interval of the Landau singularities $(0, \Lambda_{\rm Lan.}^2)$ in the plane. Spacelike QCD observables ${\cal D}(Q^2)$ are holomorphic (analytic) functions of complex $Q^2$, with the exception of the negative (timelike) semiaxis $(-\infty,-M^2_{\rm thr.})$ where $0 \leq M^2_{\rm thr.} \lesssim 1 \ {\rm GeV}^2$ is a threshold scale. The pQCD coupling $a(Q^2)$ does not reflect these properties, because of the mentioned cut interval $(0, \Lambda_{\rm Lan.}^2)$, called Landau singularities, on the positive semiaxis. Application of the Cauchy theorem to the integrand $a(Q^{' 2})/(Q^{' 2} - Q^2)$ in the complex $Q^{' 2}$-plane, with the use of the asymptotic freedom of QCD ($|a(Q^{'2})| \to 0$ when $|Q^{'2}| \to \infty$), then gives the following dispersion integral for the value of the pQCD coupling $a(Q^2)$:
\be
a(Q^2) = \frac{1}{\pi} \int_{-\Lambda_{\rm Lan.}^2- \eta}^{+\infty} d \sigma \frac{\rho_1^{\rm (pt)}(\sigma)}{(\sigma + Q^2)}, \qquad (\eta \to +0),
\label{dispa}
\ee
where $\rho_1^{\rm (pt)}(\sigma) = {\rm Im} a(Q^{' 2} = -\sigma - i \epsilon)$ is the cut discontinuity (spectral) function of $a(Q^{' 2})$. The elimination of the Landau cut contribution in this integral, while keeping the spectral function unchanged at other $\sigma$, then gives us the APT coupling \cite{ShS}
\be
\A^{\rm (APT)}(Q^2) = \frac{1}{\pi} \int_{0}^{+\infty} d \sigma \frac{\rho_1^{\rm (pt)}(\sigma)}{(\sigma + Q^2)},
\label{dispAAPT}
\ee
which has $M^2_{\rm thr.}=0$.
It is straightforward to check that the difference $\A^{\rm (APT)}(Q^2) - a(Q^2)$ at large $|Q^2| > \Lambda^2_{\rm Lan.}$ remains appreciable, $\sim (\Lambda_{\rm Lan.}^2/Q^2)$, i.e., the index $N$ in eq.~(\ref{diffAa}) is $N=1$.

The analog $\A_{\nu}^{\rm (APT)}(Q^2)$ of the power $a(Q^2)^{\nu}$ is then constructed in complete analogy, by replacing $\rho_1^{\rm (pt)}(\sigma)$ by $\rho_{\nu}^{\rm (pt)}(\sigma)={\rm Im} a(-\sigma - i \epsilon)^{\nu}$ \cite{MS96,Sh}
\be
\A_{\nu}^{\rm (APT)}(Q^2) = \frac{1}{\pi} \int_{0}^{+\infty} d \sigma \frac{\rho_{\nu}^{\rm (pt)}(\sigma)}{(\sigma + Q^2)}.
\label{dispAnuAPT}
\ee
We use in this work this form of (F)APT couplings $\A_{\nu}$ for $\nu=1,2,\ldots$, in $\MSbar$ scheme. Specifically, we apply the underlying $\MSbar$ pQCD coupling $a(Q^2;\MSbar)$ with $N_f=3$ to evaluate $\rho_n^{\rm (pt)}$ and thus $\A_n^{\rm (APT)}(Q^2)$. 

The authors of \cite{BMS05,BKS05} obtained explicit form of $\A_{\nu}^{\rm (APT)}(Q^2)$ at the one-loop level of the underlying pQCD coupling, and extended it to higher loop level by a perturbative approach \cite{BMS06,BMS10}. This theory has the name Fractional Analytic Perturbation Theory (FAPT).

Numerical programs were constructed to calculate $A_{\nu}^{\rm (APT)}(Q^2)$ up to four-loop level of the underlying $\MSbar$ pQCD coupling $a$, in \cite{NeSi,mathprg1} in Maple and/or Fortran, and in  \cite{BK,mathprg2} in Mathematica.

We note that the general approach presented in Appendix \ref{app:An} for calculation of $\A_{\nu}$ gives in the APT case (where $\rho_{\A}(\sigma) = \rho_1^{\rm (pt)}(\sigma)$, for $\sigma > 0$) approximately the same numerical results as the approach of Eq.~(\ref{dispAnuAPT}). If in the approach described in Appendix \ref{app:An} we take into account in Eq.~(\ref{AnutAnu}) [or (\ref{Ans}) when $\nu=n$ is integer] a large number of terms ${\tA}_{\nu+m}$,\footnote{This is in principle not necessary; if, for example, a physical quantity ${\cal D}(Q^2)$ is calculated up to ${\cal O}(a^4) \mapsto {\cal O}(\A_4)$, then only terms up to $\tA_4$ are in principle needed on the right-hand sides of Eqs.~(\ref{Ans}).} it turns out that the obtained $\A_{\nu}(Q^2)$ numerically converges to that in Eq.~(\ref{dispAnuAPT}). We wish to stress that the approach presented in Appendix \ref{app:An} works for any $\A$QCD, i.e., QCD with any holomorphic $\A(Q^2)$, while the approach Eq.~(\ref{dispAnuAPT}) is applicable only in the (F)APT case, i.e., when $\rho_{\A}(\sigma) = \rho_1^{\rm (pt)}(\sigma)$.

The only adjustable parameter in (F)APT (with $N_f=3$) is the scale ${\overline \Lambda}_{N_f=3}$ of the underlying QCD coupling $a(Q^2) = f(Q^2/{\overline \Lambda}_3^2)$; the high energy QCD is approximately reproduced with $N_f=3$ (F)APT when ${\overline \Lambda}_{3} = 0.45$ GeV (cf.~also Ref.~\cite{Sh}); we use this value in our analysis, but we also comment on the case ${\overline \Lambda}_3=0.40$ GeV.

\subsection{2$\delta$ and 3$\delta$ $\A$QCD}
\label{subsapp:2d3d}

This type of QCD variants \cite{2danQCD,3l3danQCD,4l3danQCD} with coupling $\A(Q^2)$ holomorphic in $Q^2 \in \mathbb{C} \backslash (-\infty, -M_{\rm thr}^2]$ are constructed on the idea that: (a) the coupling $\A(Q^2)$ at high $|Q^2| > 1 \ {\rm GeV}^2$ should practically coincide with the underlying pQCD\footnote{This means that in Eq.~(\ref{diffAa}) the index $N$ is large; as a consequence, the well demostrated success of pQCD at high momenta is reproduced in $\A$QCD.} coupling $a(Q^2)$; (b) and at moderate and low $|Q^2| \lesssim 1 {\rm GeV}^2$ the coupling should reproduce the well measured semihadronic $\tau$-lepton decay physics and possibly some other experimental indicators. We note that (F)APT does not fulfill these requirements.

 The condition (a) then implies that the spectral function $\rho_{\A}(\sigma) \equiv {\rm Im} \A(-\sigma - i \epsilon)$ is at large $\sigma > 1 \ {\rm GeV}^2$ (approximately) equal to the spectral function of the underlying pQCD, $\rho_{a}(\sigma) \equiv \rho_1^{\rm (pt)}(\sigma) \equiv {\rm Im} a(-\sigma - i \epsilon)$. At low positive $\sigma$, the unknown behavior of the spectral function $\rho_{\A}(\sigma)$ is parametrized as a sum of delta functions
\be
\rho_{\A}^{(n \delta)}(\sigma) =  \pi \sum_{j=1}^{n} {\cal F}_j \; \delta(\sigma - M_j^2)  + \Theta(\sigma - M_0^2) \rho_1^{\rm (pt)}(\sigma) \ .
\label{rhoAnd}
\ee
Implicitly, we expect $M_1^2 < M_2^2 < \cdots < M_n^2 < M_0^2$, where $M_1^2=M^2_{\rm thr.}$ is the mentioned threshold scale (expected to be $\sim m_{\pi}^2 \sim 10^{-2} \ {\rm GeV}^2$), and $M_0^2$ ($\sim 1 \ {\rm GeV}^2$) can be called the pQCD-onset scale. Application of the Cauchy theorem then gives for the coupling the expression
\bea
\A^{(n \delta)}(Q^2) \left( \equiv \frac{1}{\pi} \int_0^{\infty} d \sigma \frac{\rho_{\A}(\sigma)}{(\sigma + Q^2)} \right) & = &  \sum_{j=1}^n \frac{{\cal F}_j}{(Q^2 + M_j^2)} + \frac{1}{\pi} \int_{M_0^2}^{\infty} d \sigma \frac{ \rho_1^{\rm (pt)}(\sigma) }{(Q^2 + \sigma)} \ .
\label{AQ2}
\eea
The parametrization with $n$ delta functions in Eq.~(\ref{rhoAnd}) means that the part $\Delta \A$ of the QCD coupling originating from the unknown low-$\sigma$ part of the spectrum is parametrized by a near-diagonal Pad\'e $[n-1/n](Q^2)$ approximant
\bea
\Delta \A(Q^2)  & \equiv & \frac{1}{\pi} \int_{0}^{M_0^2} d \sigma \frac{ \rho_{\A}(\sigma) }{(Q^2 + \sigma)} \; \mapsto \; \sum_{j=1}^n \frac{{\cal F}_j}{(Q^2 + M_j^2)} = \frac{P_{n-1}(Q^2)}{Q_n(Q^2)} \equiv [n-1/n]_{\Delta \A}(Q^2),
\label{Pade}
\eea
where $P_{n-1}$ and $Q_n$ are polynomials of degree $n-1$ and $n$ in $Q^2$, respectively. If $\Delta \A$ is a Stieltjes function\footnote{If $\rho_{\A}(\sigma) \geq 0$ for $\sigma > 0$, then $\A(Q^2)$ and $\Delta \A(Q^2)$ are Stieltjes functions.}, a theorem in the Pad\'e theory \cite{BakerMorris} (cf.~also \cite{Peris}) states that there exists a sequence of Pad\'e approximants $[n-1/n](Q^2)$ which converges to $\Delta \A(Q^2)$ for any $Q^2 \in \mathbb{C} \backslash (-\infty, -M_{\rm thr}^2]$ when $n$ increases (if $\Delta \A$ is not Stieltjes, it is not known whether such a convergence is guaranteed).

The underlying pQCD coupling $a(Q^2)$ is determined, to a rather high degree of accuracy, by the world average value $a(Q^2=M_Z^2; \MSbar) = 0.1185/\pi$ \cite{PDG2014}. We use this value, and we RGE-evolve $a(\MSbar)$ to lower values of $|Q^2|$, by using the four-loop RGE (\ref{pRGE}) and three-loop quark mass threshold conditions \cite{CKS} in $\MSbar$, into the region of interest where $N_f=3$; subsequently, we change the coupling $a(Q^2; \MSbar)$ to the considered renormalization scheme to obtain the underlying coupling $a(Q^2)$; cf.~\cite{4l3danQCD} for more details. Thus we also obtain the perturbative spectral function $\rho_1^{\rm (pt)}(\sigma) = {\rm Im} a(Q^2=-\sigma - i \epsilon)$, in the considered scheme and for $N_f=3$.

At that point, the considered $\A$QCD, Eqs.~(\ref{rhoAnd})-(\ref{AQ2}) has altogether  $(2 n +1)$ parameters, namely ${\cal F}_j$ and $M^2_j$ ($j=1,\ldots, n$) and the pQCD-onset scale $M_0^2$. These are to be fixed by the conditions (a) and (b) for $\A(Q^2)$ at high and low $|Q^2|$ mentioned at the beginning of this Section \ref{subsapp:2d3d}. The condition (a) is implemented by requiring a large index value for the difference Eq.~(\ref{diffAa}); in our considered cases of 2$\delta$ and 3$\delta$ coupling we took $N=5$
\be
\A(Q^2) - a(Q^2) \sim \left( \frac{\Lambda^2}{Q^2} \right)^5.
\label{diffAaN5}
\ee
This represents altogether four conditions (for each increase of $N$ by one, from $N=1$, there is one condition).

When we take $n=2$, i.e., two delta functions in the spectral function (\ref{rhoAnd}), we have five parameters to determine; therefore one additional condition is needed. This condition will be the reproduction of the measured values of the quantity $r^{(D=0)}_{\tau}$, the semihadronic strangeless $\tau$ decay rate ratio (the leading-twist part, and with mass effects subtracted).\footnote{Higher-twist contributions in $r_{\tau}$ decay ratio are known to be strongly suppressed.} Its experimental value is approximately in the range $0.201 \pm 0.002$. Its theoretical expression can be represented as a weighted countour integral of the (massless) Adler function\footnote{This is the canonical Adler function with $N_f=3$, i.e., it is normalized in such a way that $d(Q^2)_{\rm pt} = a(Q^2) + {\cal O}(a^2)$. } $d(Q^2)$
\be
r^{(D=0)}_{\tau, {\rm th}} = \frac{1}{2 \pi} \int_{-\pi}^{+ \pi}
d \phi \ (1 + e^{i \phi})^3 (1 - e^{i \phi}) \
d(Q^2=m_{\tau}^2 e^{i \phi};D=0)  \qquad (\approx 0.201 \pm 0.001).
\label{rtaucont}
\ee
We refer to \cite{2danQCD,3l3danQCD,4l3danQCD} for details. In the $n=2$ case (2$\delta$ $\A$QCD), we still have the freedom of choosing the renormalization scheme. We took it as the Lambert scheme ($c_j = c_2^{j-1}/c_1^{j-2}$, for $j \geq 3$), with $c_2 =-4.9$ \cite{mathprg2}. It is possible to vary the value of $c_2$, but when it is different by several units from this value, either the pQCD-onset scale $M_0$ becomes appreciably higher than $\approx 1$ GeV, or the value $\A(0)$ becomes larger than one, cf.~Table 2 of \cite{mathprg2}.

When we take $n=3$, i.e., three delta functions in the spectral function (\ref{rhoAnd}), there are two additional parameters to be fixed. These two parameters are fixed by the condition $\A(Q^2) \sim Q^2$ when $Q^2 \to 0$ and the local maximum of $\A(Q^2)$ achieved at $Q^2 \approx 0.135 \ {\rm GeV}^2$. These conditions are motivated by the results of lattice calculations of the gluon and ghost dressing functions in the Landau gauge at low positive $Q^2$ \cite{LattcoupNf0} (cf.~also \cite{LattcoupNf0b}).\footnote{The calculation in \cite{LattcoupNf0,LattcoupNf0b} were performed for $N_f=0$ case. They are similar to the results when $N_f=2, 4$ \cite{LattcoupNf2,LattcoupNf4} although these results are not so precise. If the running is defined via the mentioned dressing functions, the mentioned conditions, with $\A(0)=0$, follow. Different definitions involving, in addition, a dynamical gluon mass \cite{DSEdecFreez,PTBMF}, would imply $\A(0) >0$, which holds also in 2$\delta$ $\A$QCD and in (F)APT, with the values $\A(0) = 2.0713, 1.3963$, respectively (for $N_f=3$). For a discussion on these issues, we refer to \cite{4l3danQCD}.} In our couplings we use throughout $N_f=3$, which makes them applicable in the regions $|Q^2| < 3 \ {\rm GeV}^2$. The lattice calculations were performed in the MiniMOM renormalization scheme (MM) \cite{MiniMOM,BoucaudMM,CheRet,KatMol}. Our coupling $\A(Q^2)$ was constructed in the $N_f=3$ MM scheme, but rescaled from the MM scale convention ($\Lambda_{\rm MM}$) to the usual scale convention  ($\Lambda_{\MSbar}$), the latter representing what we call Lambert MM (LMM) scheme. In \cite{3l3danQCD} we used the three-loop LMM, and in \cite{4l3danQCD} the four-loop LMM scheme. In the present work we use the latter (four-loop LMM) scheme, i.e., the coupling from \cite{4l3danQCD}.

In Table \ref{2d3dInputTab} we specify the parameters of the 2$\delta$ $\A$QCD \cite{2danQCD,mathprg2} and 3$\delta$ $\A$QCD coupling \cite{4l3danQCD}, used in the present work.
\begin{table}
  \caption{Values of the parameters of 2$\delta$ and 3$\delta$ coupling used in the present work, for $N_f=3$: the Lambert $\Lambda_{{\rm L.}}$ scale (in GeV); and the dimensionless parameters $s_j \equiv M_j^2/\Lambda_{{\rm L.}}^2$ and $f_j \equiv {\cal F}_j/\Lambda_{{\rm L.}}^2$. The ``input'' parameter choice is $\alpha_s(M_Z^2;\MSbar)=0.1185$ and $r_{\tau}^{(D=0)}=0.201$. 2$\delta$ coupling is in the Lambert scheme with $c_2=-4.9$; 3$\delta$ coupling is in the four-loop LMM scheme. We refer for details to \cite{mathprg2,4l3danQCD}.}
\label{2d3dInputTab}
\begin{ruledtabular}
\centering
\begin{tabular}{rrr|lllllllll}
 $\A$QCD & ${\overline \alpha}_s(M_Z^2)$ & $r_{\tau}^{(D=0)}$ & $s_1$ & $s_2$ & $s_3$ & $f_1$ & $f_2$ & $f_3$ & $s_0$ & $\Lambda_{{\rm L.}}$ [GeV] & $\pi \A_{\rm max}$ 
\\
\hline
2$\delta$ & 0.1185 & 0.201 & 18.734 & 1.0361 & - & 0.2929 & 0.5747 & - & 25.610 & 0.2564 & 2.0713 \\
\hline
3$\delta$ & 0.1185 & 0.201 & 3.970 & 18.495 & 474.20 & -2.8603 & 11.801 & 5.2543 & 652. & 0.11564 & 0.9156 \\
\end{tabular}
\end{ruledtabular}
\end{table}
The values of the parameters of the 2$\delta$ coupling, with $c_2=-4.9$, are slightly different from the corresponding values in \cite{mathprg2} (Table 2 there, third line), because there we used $\alpha_s(M_Z^2;\MSbar)=0.1184$ and $r_{\tau}^{(D=0)}=0.203$, while here we use $\alpha_s(M_Z^2;\MSbar)=0.1185$ and  $r_{\tau}^{(D=0)}=0.201$. We note that the maximal value of $\A(Q^2)$ for positive $Q^2$ is achieved in the 2$\delta$ case at $Q^2=0$: $\pi \A^{(2 \delta)}(0)=2.0713$; and in the 3$\delta$ case at $Q^2=Q_0^2 =0.1348 \ {\rm GeV}^2$: $\pi \A^{(3 \delta)}(Q_0^2) = 0.9156$. The pQCD-onset scales are $M_0 = \sqrt{s_0} \Lambda_{{\rm L.}} = 1.298$ GeV and $2.953$ GeV, respectively.

Other QCD running couplings $\A(Q^2)$ without Landau singularities have been constructed in the literature, some of them having at $Q^2=0$ zero value $\A(0)=0$ \cite{ArbZaits,Boucaud,mes2}, finite nonzero value (for reviews, cf.~\cite{BrazJP,Brodrev}), or infinite value \cite{Nest1}. The construction of such couplings is mostly based on the dispersive approaches similar to the ones described in this Appendix; however, such kind of dispersive approaches can also be applied to entire physical observables, cf.~\cite{ShS98,MSS2,MagrGl,mes2,DeRafael,MagrTau,Nest3a,Nest3b,NestBook}. Yet another approach is the Light Front Holographic (LFH) QCD \cite{LFH,LFHBSR}, where the coupling $\A^{(\rm (LFH})(Q^2)$ has the form $\exp(-Q^2/2 \kappa^2)$ at low positive $Q^2$, cf.~Eq.~(\ref{LFH}), and can be extended to higher positive $Q^2$ by matching of $\A(Q^2)$ and $d \A(Q^2)/d Q^2$ at a matching scale $Q_0 \sim 1 \ {\rm GeV}^2$ to pQCD \cite{LFHmatch}.

\section{Uncertainties of extracted values of $f^{p-n}_2$ and $\mu_6$: statistical and systematic}
\label{app:fit}

\begin{figure}[htb] 
\centering\includegraphics[width=100mm]{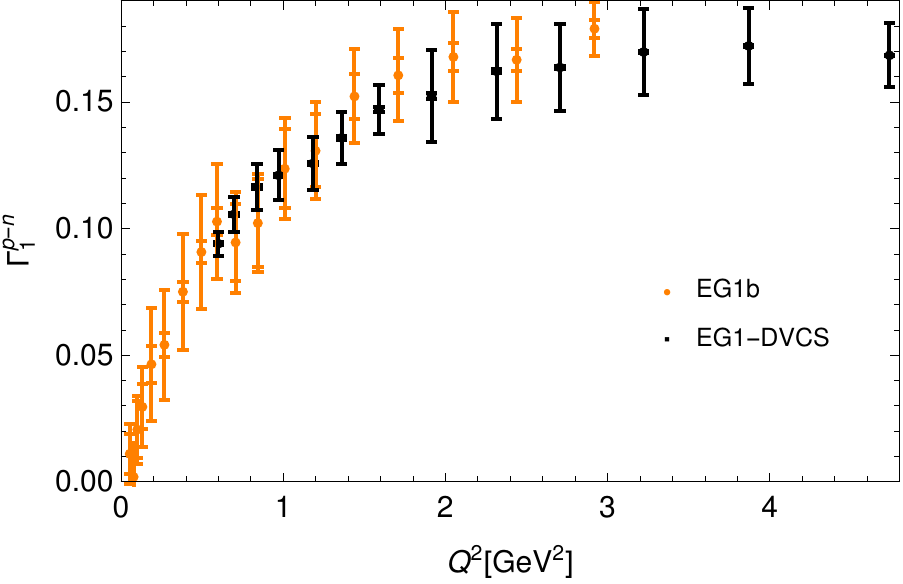}
\vspace{-0.2cm}
\caption{(color online): Experimental points $\Gamma_1^{p-n}(Q_j^2;{\rm inel})$ for two of the considered experiments: EG1b \cite{EG1b} (lighter) and EG1-DVCS \cite{DVCS} (darker). The smaller uncertainties are statistical, $\delta \Gamma_1^{p-n}(Q_j^2;{\rm inel})_{\rm stat} = \sigma_{j, {\rm stat}}$; the larger are systematic,  $\delta \Gamma_1^{p-n}(Q_j^2;{\rm inel})_{\rm sys} = \sigma_{j, {\rm sys}}$ We note that the EG1-DVCS points have very small statistical uncertainties $\sigma_{j,{\rm stat}}$ (the resolution of the Figure is too small to see them).}
\label{FigEG1bJLABN}
\end{figure}

The experimental data in the fit procedure are $\Gamma_1^{p-n}(Q_j^2;{\rm inel.})_{\rm exp.}$. The leading-twist part of the (theoretical) OPE expression (\ref{BSROPE}) has no fit parameters in our approach, apart from the renormalization scale parameter $k \equiv \mu^2/Q^2$. The latter parameter will be kept fixed at its value obtained by the approach of minimization of $\chi^2$.  We will use the following notation:
\be
z_j \equiv 1/Q_j^2, \qquad y_j\equiv \Gamma_1^{p-n}(Q_j^2;{\rm inel.})_{\rm exp.} - \Gamma_1^{p-n}(Q_j^2)^{\rm (LT)}_{\rm theor.}.
\label{fitnot}
\ee
In order to estimate the uncertainties of the extracted parameters $f^{p-n}_2(1)$ and $\mu_6$, we will consider for simplicity that not just $\mu_6$ but also $f_2^{p-n}$ does not run (i.e., that $\mu_4$ does not run) with $Q^2$. Further, we will make the following approximations: the statistical uncertainties at different points will be considered as completely uncorrelated; the systematic uncertainties at different points will be considered as completely correlated within one experiment, and completely uncorrelated between two different experiments. The systematic uncertainties $\delta \Gamma_1^{p-n}(Q_j^2;{\rm inel.})_{\rm sys}$ are in general significantly larger than the statistical uncertainties $\delta \Gamma_1^{p-n}(Q_j^2;{\rm inel.})_{\rm stat}$, cf.~Fig.~\ref{FigEG1bJLABN}.
    
\subsection{$\mu_6=0$ case}
\label{subsapp:mu60}

First we will consider the case when $\mu_6=0$ (and $f_2^{p-n}$ is not running with $Q^2$). We will estimate the uncertainty of $f_2^{p-n}$  due to the statistical uncertainties $\sigma_{j,{\rm stat}}$ at $Q_j^2$'s and then due to the systematic uncertainties $\sigma_{j, {\rm sys}}$. The renormalization scheme parameter $k$ will be kept fixed during these estimations.

In this approach, the considered $\chi^2$ is
\be
\chi^2(\mu_4) = \sum_{j} w_j ( y_j - \mu_4 z_j)^2,
\label{chi2mu4}
\ee
where we will denote
\be
w_j \equiv \frac{1}{\sigma^2_{j, {\rm stat}}},
\label{wj}
\ee
i.e., the squared statistical uncertainty at the measured point $Q^2_j$.
The minimization of $\chi^2$ then gives the extracted value of $\mu_4$
\bea
\frac{\partial \chi^2(\mu_4)}{\partial \mu_4}{\big |}_{\mu_4={\hat \mu}_4} & = & 0 \; \Rightarrow
\nonumber\\
{\hat \mu_4} & = & \frac{ \ovl{y z} }{ \ovl{z^2} },
\label{hatmu4mu60}
\eea
where we will use throughout the notation for the unnormalized ``average''
\be
{\ovl{A}} \equiv \sum_{j} w_j A(z_j) = \sum_{j} \frac{1}{ \sigma^2_{j,{\rm stat}}}{A(z_j)}.
\label{aver}
\ee
Since the deviations of $y_j \equiv y(z_j)$ due to statistical uncertainties are considered independent at different $z_j$, the rule of the sum of the squares of standard deviations is valid. It is then straightforward to deduce the square of the standard deviation for ${\hat \mu}_4$.
\be
\sigma^2({\hat \mu}_4)_{\rm stat} = \frac{1}{ \ovl{z^2} }.
\label{sigstatmu4mu60}
\ee
Then it can be checked, by Taylor expansion of $\chi^2(\mu_4)$ around the point ${\hat \mu}_4$ up to the terms $(\delta \mu_4)^2$, that the following (approximate) relation holds:
\be
\chi^2(\mu_4 = {\hat \mu}_4 \pm  \sigma({\hat \mu}_4)_{\rm stat}) =
\chi^2({\hat \mu}_4) + 1.
\label{delchi2mu60}
\ee
Any of the two relations (\ref{sigstatmu4mu60})-(\ref{delchi2mu60}) can be used to evaluate the uncertainty $(\delta {\hat \mu}_4)_{stat} \equiv \sigma({\hat \mu}_4)_{\rm stat}$. We used the relation (\ref{delchi2mu60}), which gives us in practice somewhat higher values of the uncertainty.\footnote{This difference, of a few percent ($\leq 5\%$), is presumably principally due to the effect of the running of $\mu_4(Q^2)$ with $Q^2$, the effect not accounted for in the formula (\ref{sigstatmu4mu60}).} Further, using the relation (\ref{mu4}) at $Q^2=1 \ {\rm GeV}^2$, this then gives us the uncertainty $(\delta {\hat f}_2)_{\rm stat}$\footnote{We denote, from here on, by ${\hat f}_2$ the value of $f_2^{p-n}(1{\rm GeV}^2)$ extracted by the fit procedure.}
\be
(\delta {\hat f}_2)_{\rm stat} \equiv \sigma( {\hat f}_2 )_{\rm stat} =
\frac{9}{4 M_N^2}  \sigma({\hat \mu}_4)_{\rm stat} = \frac{9}{4 M_N^2} \frac{1}{ \ovl{z^2} }.
\label{sigf2stat}
\ee 

The systematic uncertainties were estimated in the following way. For simplicity we consider for this only two experiments, namely EG1b \cite{EG1b} (experiment ``1'') and the newer JLAB EG1-DVCS results  \cite{DVCS} (experiment''2''), in the interval $0 < Q^2 < 3 \ {\rm GeV}^2$. These two experiments represent most of the available data points, and each of them covers most of the mentioned $Q^2$-interval. In each experiment, the systematic uncertainties at different points will be considered as strongly correlated, in the sense that in each experiment we will estimate the systematic uncertainty of ${\hat f}_2$ as the one obtained by averaging the deviations
\be
( \delta {\hat f}_2 )_{sys}^{(E)} \equiv \sigma({\hat f}_2)_{\rm sys}^{(E)} \approx \frac{1}{2} \left( |{\hat f}_2(UP)^{(E)} - {\hat f}_2^{(E)}|  +  |{\hat f}_2(DO)^{(E)} - {\hat f}_2^{(E)}| \right),
\label{f2sysmu60}
\ee
where $E$ is the experiment ($E=1$ or $E=2$); ``UP'' refers to the value of $f_2^{p-n}(1{\rm GeV}^2)$ extracted from the data of the experiment increased by $\sigma_{j, {\rm sys}}$ at the points $Q_j^2$ of the experiment, i.e., $\Gamma_1^{p-n}(Q_j^2;{\rm inel.}) + \sigma_{j, {\rm sys}}$; analogously, ``DO'' refers to the value extracted from   $\Gamma_1^{p-n}(Q_j^2;{\rm inel.}) - \sigma_{j, {\rm sys}}$; and ${\hat f}_2^{(E)}$ is the value of $f_2^{p-n}(1{\rm GeV}^2)$ extracted from the central points $\Gamma_1^{p-n}(Q_j^2;{\rm inel.})$ of the experiment.

Having these estimates, the question is with what relative weights to combine the two systematic uncertainties $( \delta {\hat f}_2 )_{sys}^{(E)}$ for $E=1$ and $E=2$. This will be obtained by combining the unnormalized averages of the expressions appearing in ${\hat \mu}_4^{(E)}$ of the two experiments, and assuming that the (systematic) deviations for ${\hat \mu}_4^{(E)}$ of the two experiments are mutually independent. Namely, using the definition of the (unnormalized) averages, Eq.~(\ref{aver}), for the two experiments, we have the identities
 \be
{\ovl{A}} \equiv \sum_{j} w_j A(z_j) = {\ovl{A}}^{(1)} +  {\ovl{A}}^{(2)}.
\label{aver12}
\ee
Applying these identities to ${\ovl{y z}}$ and $\ovl{z^2}$ in Eq.~(\ref{hatmu4mu60}), we obtain
\bea
{\hat \mu}_4 & = &   \frac{ \ovl{y z} }{ \ovl{z^2} } = \frac{ \ovl{y z}^{(1)} + \ovl{y z}^{(2)} }{ \ovl{z^2}^{(1)} +\ovl{z^2}^{(2)} }
\nonumber\\
& = & \alpha {\hat \mu}_4^{(1)} + (1-\alpha) {\hat \mu}_4^{(2)},
\label{mu412mu60a}
\eea
where
\be
\alpha = \frac{1}{\left( 1 + \frac{ \ovl{z^2}^{(2)} }{ \ovl{z^2}^{(1)} } \right)}.
\label{alph}
\ee
The assumption (approximation) that the two systematic deviations of ${\hat \mu}_4$ of experiments 1 and 2 are independent, then leads us to the standard deviation $(\delta {\hat \mu}_4)_{sys}$
\be
 (\delta {\hat \mu}_4)_{sys} \equiv \sigma({\hat \mu}_4)_{\rm sys} = \left[ \alpha^2  \sigma^2({\hat \mu}_4)_{\rm sys}^{(1)}  + (1 - \alpha)^2 \sigma^2({\hat \mu}_4)_{\rm sys}^{(2)} \right]^{1/2},
\label{mu412mu60b}
\ee
and by relation (\ref{mu4}) analogously 
\be
 ( \delta {\hat f}_2 )_{sys} \equiv \sigma({\hat f}_2)_{\rm sys} = \left[ \alpha^2  \sigma^2({\hat f}_2)_{\rm sys}^{(1)}  + (1 - \alpha)^2 \sigma^2({\hat f}_2)_{\rm sys}^{(2)} \right]^{1/2},
\label{f212mu60b}
\ee
where the estimates  $\sigma({\hat f}_2)^{(E)}_{\rm sys}$ ($E=1,2$) are given in Eq.~(\ref{f2sysmu60}). Eq.~(\ref{f212mu60b}) represents thus an estimate of the uncertainty of the extracted value of $f_2^{p-n}(1 {\rm GeV}^2)$ due to systematic uncertainties of the experimental data. In practice, it turns out that this uncertainty is dominated by the results  of the experiment 2 (JLAB EG1-DVCS) \cite{DVCS}, i.e., $(1 - \alpha) \approx 1$. This is so because $\ovl{z^2}^{(2)} \gg \ovl{z^2}^{(1)}$, since the experiment 2 has significantly larger values of $w_j$, i.e., significantly smaller values of $\sigma^2_{j, {\rm stat}}$.

\subsection{$\mu_6 \not= 0$ case}
\label{subsapp:mu6not0}

When the coefficient $\mu_6$ is included in the truncated OPE (\ref{BSROPE}) as a fit parameter, the analysis is analogous to the previous Sec.~\ref{subsapp:mu60}, except that now the algebra is more involved. The values of ${\hat \mu}_4$ and ${\hat \mu}_6$ are obtained by simultaneous minimization of 
\be
\chi^2(\mu_4,\mu_6) = \sum_{j} w_j ( y_j - \mu_4 z_j - \mu_6 z_j^2)^2,
\label{chi2mu4mu6}
\ee    
with respect to $\mu_4$ and $\mu_6$. This gives
\be
\label{hatmu4mu6}
{\hat \mu_4}  =  \frac{ -\ovl{y z^2} \ \ovl{z^3} + \ovl{y z} \ \ovl{z^4} }{D},
\qquad
{\hat \mu_6}  =  \frac{ \ovl{y z^2} \ \ovl{z^2} - \ovl{y z} \ \ovl{z^3} }{D},
\ee
where
\be
D \equiv ( \ovl{z^2} \ \ovl{z^4} - \ovl{z^3} \ \ovl{z^3} ).
\label{Dnot}
\ee
The corresponding squares of the standard deviations are
\be
\sigma^2({\hat \mu}_4)_{\rm stat} = \frac{ \ovl{z^4} }{ D}, \qquad
\sigma^2({\hat \mu}_6)_{\rm stat} = \frac{ \ovl{z^2} }{ D}.
\label{sigstatmu4mu6}
\ee
When using Taylor expansion of $\chi^2(\mu_4,\mu_6)$ around the point $({\hat \mu}_4,{\hat \mu}_6)$ up to the terms quadratic in the deviations, it can be checked that the following (approximate) relations hold:
\bes
\label{delchi2}
\bea
\chi^2(\mu_4 = {\hat \mu}_4 \pm  \sigma({\hat \mu}_4)_{\rm stat}, \mu_6={\hat \mu}_6) &=&
\chi^2_{\rm min} + \frac{ \ovl{z^2} \ \ovl{z^4} }{ D }
\label{delchi2mu4}
\\
\chi^2( \mu_4={\hat \mu}_4, \mu_6 = {\hat \mu}_6 \pm  \sigma({\hat \mu}_6)_{\rm stat}) &=&
\chi^2_{\rm min} + \frac{ \ovl{z^2} \ \ovl{z^4} }{ D }
\label{delchi2mu6}
\eea
\ees 
We determined the values $\sigma({\hat \mu}_4)_{\rm stat}$ and $\sigma({\hat \mu}_6)_{\rm stat}$ from Eqs.~(\ref{delchi2}).\footnote{Eq.~(\ref{delchi2mu6}) and Eqs.~(\ref{sigstatmu4mu6}) give the same result for $\sigma({\hat \mu}_6)_{\rm stat}$. For $\sigma({\hat \mu}_4)_{\rm stat}$ [and $\sigma({\hat f}_2)_{\rm stat}$, cf.~Eq.~(\ref{sigf2stat})], the result of Eq.~(\ref{delchi2mu4}) differs from that of Eqs.~(\ref{sigstatmu4mu6}) by a few percent ($\leq 5 \%$), principally due the the effect of the running of $f_2^{p-n}(Q^2)$ with $Q^2$.}

Also the systematic uncertainties were estimated analogously to the case of $\mu_6=0$, cf.~Sec.~\ref{subsapp:mu60}, only the algebra is now more involved. The basis is again the identity (\ref{aver12}), but this time for the quantities $\ovl{y z^2}$ and $\ovl{y z}$ which appear in ${\hat \mu}_4$ and ${\hat \mu}_6$, Eqs.~(\ref{hatmu4mu6}) and can thus be expressed by ${\hat \mu}_4$ and ${\hat \mu}_6$
\be
\label{yzyz2}
\ovl{y z}  =  {\hat \mu}_4 \ovl{z^2} + {\hat \mu}_6 \ovl{z^3},
\qquad
\ovl{y z^2}  =  {\hat \mu}_4 \ovl{z^3} + {\hat \mu}_6 \ovl{z^4}.
\ee
Using this, we obtain
\be
\label{hatmu12}
{\hat \mu}_4 =  {\widetilde \mu}_4^{(1)} + {\widetilde \mu}_4^{(2)},
\qquad
{\hat \mu}_6 =  {\widetilde \mu}_6^{(1)} + {\widetilde \mu}_6^{(2)},
\ee
where 
\bes
\label{wtmu}
\bea
 {\widetilde \mu}_4^{(1)} &=& {\widetilde \alpha} {\hat \mu}_4^{(1)} - \kappa_{34} {\hat \mu}_6^{(1)},
\qquad
 {\widetilde \mu}_4^{(2)} = (1-{\widetilde \alpha}) {\hat \mu}_4^{(2)} + \kappa_{34} {\hat \mu}_6^{(2)},
\label{wtmu4}
\\
 {\widetilde \mu}_6^{(1)} &=& {\widetilde \beta} {\hat \mu}_6^{(1)} + \kappa_{23} {\hat \mu}_4^{(1)},
\qquad
 {\widetilde \mu}_6^{(2)} = (1-{\widetilde \beta}) {\hat \mu}_6^{(2)} - \kappa_{23} {\hat \mu}_4^{(2)}.
\label{wtmu6}
\eea
\ees
As earlier, we denote by '(1)' and '(2)' the experiments 1 (EG1b, \cite{EG1b}) and 2 (EG1-DVCS, \cite{DVCS}), and the constants appearing in Eqs.~(\ref{wtmu}) are
\bes
\label{wtnot}
\bea
\kappa_{34} & = & \frac{1}{D^{(1+2)}} \left( - \ovl{z^3}^{(1)} \ovl{z^4}^{(2)} + \ovl{z^3}^{(2)} \ovl{z^4}^{(1)} \right),
\\
\kappa_{23} & = & \frac{1}{D^{(1+2)}} \left( - \ovl{z^2}^{(1)} \ovl{z^3}^{(2)} + \ovl{z^2}^{(2)} \ovl{z^3}^{(1)} \right),
\\
{\widetilde \alpha} & = & \frac{1}{D^{(1+2)}} \left( D^{(11)} + D^{(12)} \right),
\\
{\widetilde \beta} & = & \frac{1}{D^{(1+2)}} \left( D^{(11)} + D^{(21)} \right),
\eea
\ees
where the $D$-terms are defined as
\bes
\label{Ds}
\bea
D^{(ij)} & = & \ovl{z^2}^{(i)} \ovl{z^4}^{(j)}- \ovl{z^3}^{(i)} \ovl{z^3}^{(j)} \quad (i,j=1,2),
\\
D^{(1+2)} &=& \sum_{j=1}^2 \sum_{i=1}^2 D^{(ij)} = \ovl{z^2} \ \ovl{z^4}- \ovl{z^3} \ \ovl{z^3},
\eea
\ees
where in the last expression on the right-hand side, the unnormalized averages are made over the experiments 1 and 2.\footnote{This is somewhat different from the unnormalized averages $\ovl{A}$ appearing in Eqs.~(\ref{hatmu4mu6})-(\ref{sigstatmu4mu6}) which are over all the experimental points of the fit, i.e., over more than two experiments.}

The estimates of the systematic uncertainties for the quantities ${\widetilde \mu}_4^{(E)}$ and ${\widetilde \mu}_6^{(E)}$, for experiments $E=1, 2$, are evaluated in complete analogy with Eq.~(\ref{f2sysmu60}) of the previous Sec.~\ref{subsapp:mu6not0}
\bea
\label{wtmusys}
\left( \delta {\widetilde \mu}_N \right)_{sys}^{(E)} \equiv {\sigma}( {\widetilde \mu}_N)_{\rm sys}^{(E)} & \approx &  \frac{1}{2} \left(  |{\widetilde \mu}_N(UP)^{(E)} - {\widetilde \mu}_N^{(E)}|  +  |{\widetilde \mu}_N(DO)^{(E)} - {\widetilde \mu}_N^{(E)}| \right),
\eea
where $N=4$ or $N=6$. For explanations of the notation 'UP' and 'DO' we refer to the previous Sec.\ref{subsapp:mu6not0}. In our approximation, we consider that the systematic uncertainties in the experiments 1 and 2 are mutually independent; this, in conjunction with the relations (\ref{hatmu12}), then gives 
\bes
\label{sigmu4mu612}
\bea
(\delta {\hat \mu}_4)_{sys} \equiv \sigma({\hat \mu}_4)_{\rm sys} & = &
\left[  {\sigma}^2({\widetilde \mu}_4^{(1)})_{\rm sys} + 
{\sigma}^2({\widetilde \mu}_4^{(2)})_{\rm sys} \right]^{1/2},
\label{sigmu412}
\\
(\delta {\hat \mu}_6)_{sys} \equiv \sigma({\hat \mu}_6)_{\rm sys} & = &
\left[  {\sigma}^2({\widetilde \mu}_6^{(1)})_{\rm sys} + 
{\sigma}^2({\widetilde \mu}_6^{(2)})_{\rm sys} \right]^{1/2},
\label{sigmu612}
\eea
\ees
The systematic uncertainty of the extracted value $f_2^{p-n}(1{\rm GeV}^2)$ is obtained then from Eq.~(\ref{sigmu412}) via the relation (\ref{mu4})
\bea
( \delta {\hat f}_2 )_{sys} \equiv \sigma({\hat f}_2)_{\rm sys} & = & \frac{9}{4 M_N^2} \sigma({\hat \mu}_4)_{\rm sys}.
\label{sigf212}
\eea

\subsection{``Massive'' OPE case}
\label{subsapp:mass}

When the truncated OPE has, instead, a ``massive'' term Eq.~(\ref{BSROPEM}), a similar estimation of the statistical and systematic uncertainties of the extracted parameters $f_2^{p-n}(1{\rm GeV}^2)$ and $M^2$ can be made. This is so because, when expanding the ``massive'' higher-twist term in powers of $1/Q^2$, we obtain
\be
\mu_6(M^2) = - M^2 \mu_4; \qquad  M^2 = -\frac{\mu_6}{\mu_4}.
\label{mu6M2}
\ee
In the first approximation, the ``massive'' case can thus be considered as the case of truncated OPE (\ref{BSROPE}). To estimate the statistical uncertainties of the extracted $f_2^{p-n}(1{\rm GeV}^2)$ and $M^2$, we decided to apply the relations of the type (\ref{delchi2})
\bes
\label{delchi2M}
\bea
\chi^2(\mu_4 = {\hat \mu}_4 \pm  \sigma({\hat \mu}_4)_{\rm stat}, M^2={\hat M}^2) &=&
\chi^2_{\rm min} + \frac{ \ovl{z^2} \ \ovl{z^4} }{ D }
\label{delchi2mu4M}
\\
\chi^2( \mu_4={\hat \mu}_4, M^2 = {\hat M}^2 \pm  \sigma({\hat M}^2)_{\rm stat}) &=&
\chi^2_{\rm min} + \frac{ \ovl{z^2} \ \ovl{z^4} }{ D }.
\label{delchi2mu6M}
\eea
\ees 

To estimate the systematic uncertainties in the ``massive'' case, the relations (\ref{sigmu4mu612}) for $\mu_4$ and $\mu_6$ were used. Namely, differentiation of the relation (\ref{mu6M2}) gives an approximate relation
\bea
\sigma^2 (M^2)_{\rm sys} & \approx & \left( \frac{{\hat \mu}_6}{{\hat \mu}_4^2} \right)^2 \sigma^2({\hat \mu}_4)_{\rm sys} + \frac{1}{{\hat \mu}_4^2} \sigma^2({\hat \mu}_6)_{\rm sys} - 2 \left( \frac{ {\hat \mu}_6}{{\hat \mu}_4^3} \right) \langle \delta {\hat \mu}_4 \delta {\hat \mu}_6 \rangle_{sys},
\label{sigM2sys}
\eea
where in the first two terms on the right-hand side we can use the expressions (\ref{sigmu4mu612}), and in the last term the correlator $\langle \delta {\hat \mu}_4 \delta {\hat \mu}_6 \rangle_{sys}$ can be estimated in a similar way as other correlators described above. Motivated by the relations (\ref{hatmu12}), we chose the following estimate for this correlator:
\bea
\langle \delta {\hat \mu}_4 \delta {\hat \mu}_6 \rangle_{sys} & = &
\frac{1}{2} \sum_{E=1}^2 {\Big [} \left({\widetilde \mu}_4^{(E)}(UP) -  {\widetilde \mu}_4^{(E)}\right)  \left({\widetilde \mu}_6^{(E)}(UP) -  {\widetilde \mu}_6^{(E)}\right) 
\nonumber\\
&& + \left({\widetilde \mu}_4^{(E)}(DO) -  {\widetilde \mu}_4^{(E)}\right)  \left({\widetilde \mu}_6^{(E)}(DO) -  {\widetilde \mu}_6^{(E)}\right) {\Big ]}.
\label{mu4mu6M}
\eea
We note that in the ``massive'' case the systematic uncertainties are often quite large, so the estimate (\ref{sigM2sys}) should be regarded often as only a rough approximation.

\section{Charm mass effects in BSR}
\label{app:Nf34}

The contributions of the finite charm quark mass, i.e., the effects beyond $N_f=3$, in the polarized BSR at NLO were evaluated in \cite{BFF}. When ignoring the $b$-quark contributions (considering $m_b \to \infty$), the mentioned effects at NLO can be expressed with the function $C_{\rm pBJ}^{\rm mass.,(2)}(\xi_c)$, where $\xi_c \equiv Q^2/m_c^2$ which is assumed in \cite{BFF} to be $\xi_c \gtrsim 1$, and $m_c \approx 1.59$ GeV is the pole mass. This function appears in the NLO coefficient (at $a^2$) when the perturbation expansion (\ref{DBSpt}) is reexpressed in terms of the $N_f=4$ coupling $a_{N_f=4}$
\bea
{\cal D}_{\rm BS}(Q^2)_{\rm pt} &=& a(Q^2)_{N_f=4} + a(Q^2)^2_{N_f=4}
\left\{ \frac{55}{12} - \frac{1}{3} \left[ N_f -1 + C_{\rm pBJ}^{\rm mass.,(2)}(\xi_c) \right] \right\} + {\cal O}(a^3),
\label{DBSmc1}
\eea
where $N_f=4$ and
\bea
C_{\rm pBJ}^{\rm mass.,(2)}(\xi) & = & \frac{1}{2520} {\bigg \{} \frac{1}{\xi} ( 6 \xi^2 + 2735 \xi + 11724 ) - \frac{\sqrt{\xi+4}}{\xi^{3/2}} ( 3 \xi^3 + 106 \xi^2 + 1054 \xi + 4812) \ln \left[ \frac{ \sqrt{\xi + 4} + \sqrt{\xi} }{\sqrt{\xi + 4} - \sqrt{\xi} } \right]
\nonumber\\
&&
- 2100 \frac{1}{\xi^2} \ln^2 \left[ \frac{ \sqrt{\xi + 4} + \sqrt{\xi} }{\sqrt{\xi + 4} - \sqrt{\xi} } \right] + (3 \xi^2 + 112 \xi + 1260) \ln \xi {\bigg \}} \qquad (\xi \gtrsim 1).
\label{CBj1}
\eea
When $Q^2 \gg m_c^2$ ($\xi \gg 1$), this function approaches unity quite slowly
\be
C_{\rm pBJ}^{\rm mass.,(2)}(\xi)  = 1 - \frac{8}{3} \frac{\ln \xi}{\xi} + \frac{34}{9\xi }  + {\cal O}\left(  \frac{\ln^2 \xi}{\xi^2} \right).
\label{CBj2}
\ee
In this limit ($\xi_c \to \infty$) this then gives in the expansion (\ref{DBSmc1}) the massless expression for the NLO coefficient $d_1(N_f)$ with $N_f=4$
\be
d_1(N_f)= \frac{55}{12} - \frac{1}{3} N_f.
\label{e2NS}
\ee
As noted, the convergence to the $N_f=4$ case (four massless quarks) is rather slow in BSR. Specifically, $3+C_{\rm pBJ}^{\rm mass.,(2)}(Q^2/m_c^2) \approx 3.13$, $3.36$, $3.73$, $3.83$ for $Q^2=5$, $10$, $50$, $100 \ {\rm GeV}^2$, respectively. This indicates that at $Q^2=5 \ {\rm GeV}^2$ (the highest considered experimental $Q^2$ in BSR) we are still rather far away from the $N_f=4$ case. Therefore, it appears reasonable to use $N_f=3$ (i.e., decoupled charm quark) in the polarized BSR for the interval $0 < Q^2 < 5 \ {\rm GeV}^2$ considered in the present work.

\end{document}